\documentclass[12pt]{article}
\usepackage{etex}
\usepackage{epsfig}
\usepackage{epsf}
\usepackage{latexsym}
\usepackage{amsfonts,amssymb,amsmath} 
\usepackage{mathtools}
\usepackage{color}
\usepackage[a4paper,vmargin=3cm,hmargin=2.1cm]{geometry}
\usepackage{tikz}
\usepackage[all]{xy}
\epsfverbosetrue
\newcommand{\goth}[1]{\mathfrak{#1}}
\newcommand{\double}[1]{\mathbb{#1}}

\newcommand{\rr}{\double{R}}

\newcommand{\tg}{\goth{t}}

\newcommand{\hh}{\mathcal{H}}

\newcommand{\de}{\hbox{\rm{d}}}

\newcommand{\pa}{\partial}

\newcommand{\lb}{\left[}
\newcommand{\rb}{\right]}
\newcommand{\lp}{\left(}
\newcommand{\rp}{\right)}
\newcommand{\la}{\left\{}
\newcommand{\ra}{\right\}}
\newcommand{\ul}[1]{\underline{#1}}

\newcommand{\dpp}{\vcentcolon}
\newcommand{\bb}{\begin{eqnarray}}
\newcommand{\ee}{\end{eqnarray}}
\newcommand{\eee}{\nonumber\end{eqnarray}}
\newcommand{\qq}{\quad}

\newcommand{\nc}{\newcommand}
 \nc{\alf}{\alpha} %\nc{\la}{\lambda} \nc{\La}{\Lambda}
\nc{\ze}{\zeta}
\nc{\tht}{\theta} \nc{\T}{\Theta} \nc{\be}{\beta}  \nc{\eps}{\epsilon} 
\nc{\ga}{\gamma}  \nc{\De}{\Delta} 
 \nc{\G}{\Gamma}  \nc{\vphi}{\varphi}
 \nc{\si}{\sigma}  \nc{\ka}{\kappa}   \nc{\Si}{\Sigma} 
\nc{\om}{\omega}  \nc{\chic}{\widehat{\chi}}

\nc{\qqq}{\quad\quad}               

 \nc{\Om}{\Omega}
\nc{\nf}{\infty}   \nc{\dl}{\mathop{\smash{\cal L}}}  \nc{\black}{\rule{3mm}{3mm}}
\nc{\ol}{\overline}        \nc{\und}{\underline} 
\nc{\beq}{\begin{equation}}  \nc{\eeq}{\end{equation}}  \nc{\pt}{\partial}  
   \nc{\dst}{\displaystyle}  \nc{\na}{\nabla} 
\nc{\nnb}{\nonumber}    \nc{\bs}{\backslash}        \nc{\mb}{\mathbb}   
\nc{\sn}{{\rm sn}\,} \nc{\cn}{{\rm cn}\,}     \nc{\dn}{{\rm dn}\,} \nc{\nin}{\noindent}
\nc{\ti}{\tilde}   \nc{\wti}{\widetilde}   \nc{\h}{\hat}  \nc{\wh}{\widehat}
\nc{\tpsi}{\wti{\psi}}   \nc{\tphi}{\wti{\phi}}  \nc{\tH}{\wti{H}} 

\newcounter{muni}
\newenvironment{remunerate}{\begin{list}{{\rm \arabic{muni}.}}
{\usecounter{muni}
\setlength{\leftmargin}{0pt}\setlength{\itemindent}{38pt}}}{\end{list}}

\nc{\brm}{\begin{remunerate}}   \nc{\erm}{\end{remunerate}}

\nc{\stg}{\mathop{\smash{*}}}
\nc{\st}{\mathop{\smash{\delta}}}
\nc{\barr}{\begin{array}}   \nc{\earr}{\end{array}}   \nc{\dg}{\dagger}
\nc{\mtvb}{\mathversion{bold}}   \nc{\mtvn}{\mathversion{normal}}

\begin{document}

\thispagestyle{empty}

\begin{center}
${}$
\vspace{3cm}

{\Large\textbf{Axial Bianchi IX and its Lema{\^i}tre-Hubble diagram}} \\

\vspace{2cm}

{\large
Galliano Valent\footnote{Aix Marseille Univ, Universit\'e de Toulon, CNRS, CPT, IPhU, Marseille, France;
\\\indent\qq
on leave from LPMP, Aix-en-Provence, France
\\\indent\qq
galliano.valent@orange.fr},
Andr\'e Tilquin\footnote{Aix Marseille Univ,  CNRS/IN2P3, CPPM, IPhU, Marseille, France \\\indent\qq tilquin@cppm.in2p3.fr },
Thomas Sch\"ucker\footnote{
Aix Marseille Univ, Universit\'e de Toulon, CNRS, CPT, IPhU, Marseille, France\\\indent\qq
thomas.schucker@gmail.com }
}

\vspace{3cm}

{\large\textbf{Abstract}}
\end{center}

We compute the Lema{\^i}tre-Hubble diagram for axial Bianchi IX universes with comoving dust. We motivate our choice by defining a {\it minimal} symmetry breaking of the cosmological principle. This criterium admits only two possibilities: the axial Bianchi I and IX universes. The latter have positive curvatures and reduce to the former in the zero curvature limit. Remarkably, negative curvatures are excluded by this minimal symmetry breaking in presence of comoving dust, as shown by Farnsworth in 1967. We present an alternative proof of his result.

\vspace{2cm}

\hfill{\em To the memory of Vaughan Jones}

\vspace{2cm}


\section{Introduction}

Symmetries reduce the complexity of systems and may lead to useful approximations. Observed deviations from the symmetric approximation may destroy all symmetries or only some. 

Take the surface of the Earth. In good approximation, this surface is maximally symmetric, i.e. a sphere with its 3-dimensional isometry group. Easily observed deviations, like Mount Everest, are of the order of one per mille, $8.8\, {\rm km}\cdot 2\pi \ / (40\,000\,{\rm km})\,\approx\, 1.4\cdot10^{-3}.$ They destroy all symmetries; let us call these deviations geographic. There are other deviations from maximal symmetry, of the order of three per mille, but more difficult to observe, that respect one of the three isometries. We call them geometric.  Indeed these deviations admit a simple geometric description, in terms of an oblate ellipsoid. Our polar radius is about 21.3 km shorter than the equatorial ones, $21.3\, {\rm km}\cdot 2\pi /  (\,40\,000\,{\rm km})\,\approx\, 3.3\cdot10^{-3}.$

In relativistic cosmology, one is tempted to start with a maximally symmetric space-time. A fundamental theorem of pseudo-Riemannian geometry states that the isometry group of a $d$-dimensional space or space-time is of dimension less than or equal to $d\,(d+1)/2$. For the surface of the Earth, $d=2$, this maximal dimension is indeed three; for our space-time, $d=4$, the maximal dimension is 10 and there are three families of maximally symmetric space-times: anti de Sitter spaces, Minkowski space and de Sitter spaces. However these three families do not support dynamics and cosmology went for less symmetry by adopting the {\it cosmological principle}. It postulates that space-time has maximally symmetric 3-spaces of simultaneity. Again they come in three families: pseudo-spheres, Euclidean space and spheres. By the above theorem we therefore start our approximation with 6-dimensional isometry groups only: $O(1,3)$,  $O(3)\ltimes\rr^3$  and $O(4)$. Space-times satisfying the cosmological principle are named after Friedman-Lema{\^i}tre-Robertson-Walker or simply Robertson-Walker.

The Cosmic Microwave Background data does show significant deviations from this approximation and they are of the order of $10^{-5}$. 
Bianchi universes close to Robertson-Walker universes (I, V, VII$_h$, IX) have been used to try and decipher a 3-dimensional subgroup of the 6-dimensional isometry groups in these data \cite{EMC}-\cite{cea4}; a reader-friendly overview by Pereira  \& Pitrou \cite{pp} also contains more references.
It is fair to say that we still do not know whether the Cosmic Microwave Background data does contain geometric deviations. 

In 1898 Luigi Bianchi \cite{Bianchi} classifies all real 3-dimensional Lie algebras. To situate his work we note that there is only one 1-dimensional Lie algebra: the Abelian one. There are two 2-dimensional Lie algebras: the Abelian one and the (solvable) one of $2 \times 2$ triangular matrices with vanishing trace. Bianchi's list starts with seven  3-dimensional Lie algebras: Bianchi~I, II, III, IV, V, VIII and IX. Bianchi~I is the Abelian algebra, Bianchi~II is the Heisenberg algebra, Bianchi~III is the direct sum of the 1-dimensional and the solvable 2-dimensional Lie algebras. 
The last, Bianchi IX, is $so(3)\cong su(2)$, the Lie algebra of the rotation group in $\rr^3$ or its spin lift.  
In addition Bianchi finds two uncountable  families of 3-dimensional Lie algebras, each indexed by one real parameter $h$: Bianchi~VI$_h$ with $h\not=0,\,h\not=1$, and VII$_h$ with $h\geq 0$. Note that Bianchi~{VI$_0$} is isomorphic to Bianchi~{III} and Bianchi~{VI$_1$} is isomorphic to Bianchi~{V}, these two are special and do deserve different names. Note also that Bianchi~{VII$_{-h}$} is isomorphic to Bianchi~{VII$_{+h}$}.

As early as 1933, Georges Lema{\^i}tre \cite{lemaitre}, following a recommandation by Albert Einstein, considers Bianchi I universes without using this name.

Besides via observation of the Cosmic Microwave Background, deviations from the cosmological principle may also be observed via 
drifts in redshift and position of galaxies and quasars \cite{cospar}-\cite{mppuDrift}, via privileged directions in the Lema{\^i}tre-Hubble diagram of type 1a supernovae \cite{hemi1}-\cite{Krish} 
 and of quasars and gamma-ray bursts \cite{Secr}, \cite{Luon},
 via weak lensing data \cite{ppuWeakL},
 via galaxy cluster scaling relations \cite{Mig}, \cite{Mig2},
 via radio galaxies \cite{siew}
  and perhaps via black hole or neutron star mergers in the future \cite{cup}-\cite{Colgain}.

In a few years the Vera Rubin Observatory (previously referred to as the Large Synoptic Survey Telescope, LSST) will start improving the precision of the Lema{\^i}tre-Hubble diagram by one to two orders of magnitude~\cite{stv}.

So far only Bianchi~I universes have been used to decipher geometrical deviations in the Lema{\^i}tre-Hubble diagram \cite{stv}-\cite{fpu} and we would like to address two questions:
\begin{itemize}\item
 Why {\it three}-dimensional subgroups of the six-dimensional isometry groups? 
 \item
 What about the other Bianchi universes for modeling geometric deviations in the Lema{\^i}tre-Hubble diagram? 
\end{itemize}

For the second question we will use Einstein's equations with a cosmological constant. We assume that radiation can be neglected and that galaxies, supernovae and dark matter are modeled by comoving dust.  `Comoving'  means that there is a  system of coordinates with respect to which the energy-momentum tensor has vanishing time-space components and  `dust' means that the space-space components vanish.

 There is an alternative way of weakening the cosmological principle: consider multiply connected universes with six Killing vectors but less global symmetries. This approach has excellent reviews by Lachi\`eze-Rey \& Luminet \cite{lalu,lum}.

We close  the introduction by mentioning that a short version of the present work is also available \cite{short}. 

 \section{Counting Killing vectors \label{sect counting}}
 
In his 1898 work \cite{Bianchi}, Bianchi  not only classifies the real 3-dimensional Lie algebras but also their {\it local} representations as Killing vectors on spaces of dimension $d=3$. By systematically solving the Killing equations, he also computes all possible representations of Lie algebras of dimension $n=4$ on spaces of dimension three. He proves that the 4-dimensional Lie algebras admitting such representations on 3-spaces necessarily have three-dimensional Lie subalgebras. He goes on to prove that Lie algebras of dimension $n=5$ cannot be represented on 3-spaces. 

Five years later his student Guido Fubini \cite{Fubini} generalizes this last result to the {\it global} setting of manifolds and groups  and to higher dimensions: 
\\[2mm]
 {\bf Theorem} (Fubini 1903): Consider a Riemannian manifold of dimension $d\ge 3$. Its isometry group cannot be of dimension $n=d\,(d+1)/2-1$.\\[2mm]
The $d=2$ dimensional cylinder 
is multiply connected.
It has two global isometries, the translations, and one only local isometry, the rotation, and it is a counter-example to Fubini's theorem.

Let us come back to cosmology. Denoting by $n$ the dimension of the isometry group of the 3-spaces of simultaneity, let us take $\Delta \dpp=6-n$ to measure the extent of symmetry breaking of the cosmological principle. The minimal value $\Delta=0$ means no breaking and the maximal value $\Delta =6$ means geographic breaking. By  Fubini's theorem \cite{Fubini} the next to minimal value $\Delta =1$ is impossible. 
 
 Therefore the minimal symmetry breaking has $\Delta =2 $ i.e. four Killing vectors. 
 The cos\-mological principle produces  approximate descriptions of observed data with a low number of parameters. With increasing experimental precision, tensions arise and we may resort to symmetry breaking in order to add more parameters. However here parsimony is called for and minimal symmetry breaking is a wise choice.

  Since observations are in good agreement with the cosmological principle, we add to our definition of minimal symmetry breaking that the metric be  `close' to Robertson-Walker. There are several criteria for convergence, for simplicity let us choose uniform convergence.
 The simplest example of minimal symmetry breaking is Bianchi~I with an additional axial symmetry:
  \bb \de \tau^2 = \de t^2 - a(t)^2\,[\de x^2
 +\de y^2]- c(t)^2\,\de z^2,\label{bianchi1}\ee
 and its four Killing vectors:
 \bb \,\frac{\pa}{\pa x}\,,\,\frac{\pa}{\pa y}\,,\,\frac{\pa}{\pa z}\,,\qq \text{and}\qq x\,\frac{\pa}{\pa y}-y\,\frac{\pa}{\pa x}\,.
 \ee
 In the limit $c-a\rightarrow 0$, it is as close to flat Robertson-Walker as you like.

In Bianchi's classification there are only two more examples of minimal symmetry breaking in the above sense: the axial Bianchi~V with negative curvature and the axial Bianchi~IX with positive curvature. The latter is also known as Berger sphere \cite{berger}. 
Note that some authors, e.g. \cite{KingEllis}, use the adjective `locally rotationally symmetric  (L.R.S.)' instead of axial.

For completeness we mention the family Bianchi~{VII$_h$}\,. It  also contains Robertson-Walker:  the flat one for $h=0$ and the one with negative curvature for positive $h$. However their axial versions are identical to axial Bianchi~I for $h=0$ and to axial Bianchi~V for $h>0$.

We close this section by sketching a proof of the well known fact:
  the Bianchi universes II, III, IV, VI$_h\,,$ VIII, do not contain any  Robertson-Walker universe. To this end we may restrict our attention to the 3-dimensional spaces of simultaneity on which the 3-dimensional Lie algebras of Killing vectors are represented. These spaces have been exhaustively analyzed by Bianchi \cite{Bianchi}. We also use the theorem valid in three dimensions: a metric has maximal local symmetry if and only if it is `Einstein', i.e. its Ricci tensor is proportional to its metric tensor. 

Thanks to the three symmetries on the Bianchi 3-spaces,  their Ricci tensors can be computed purely algebraically by using orthonormal frames related to the Maurer-Cartan forms $\sigma ^i,\,i=1,2,3$. Here are these forms in coordinates $x^1=\dpp x,\,x^2=\dpp y,\,x^3=\dpp z$, which are close to Terzis' \cite{ter}:
\bb 
\text{I}&
\de x,\,\de y,\,\de z&\nonumber\\
\text{II}&
\de x+y\,\de z,\,\de y,\,\de z&\nonumber\\
\text{III}&
e^{-z}\de x,\de y,\,\de z
&\nonumber\\
\text{IV}&
e^{-z}(\de x-z\,\de y),\,e^{-z}\de y,\,\de z
&\nonumber\\
\text{V}&
e^{-z}\de x,\,e^{-z}\de y,\,\de z
&\nonumber\\
\text{VI}_h&
e^{-z}\de x,\,e^{-hz}\de y,\,\de z
&\nonumber\\
\text{VII}_h&
e^{-hz}(\cos z\,\de x-\sin z\,\de y),\,\,e^{-hz}(\sin z\,\de x+\cos z\,\de y),\,\de z
&\nonumber\\
\text{VIII}&\cos z\,\de x-\cosh x\, \sin z\,\de y,\,\sin z\,\de x+\cosh x\, \cos z\,\de y,\,-\sinh x\,\de y+\de z
\nonumber\\
\text{IX}&
\cos z\,\de x-\cos x\, \sin z\,\de y,\,\sin z\,\de x+\cos x\, \cos z\,\de y,\,\sin x\,\de y+\de z
\nonumber
\ee
Now the line elements read
\bb \de s^2 = (a_1 \sigma ^1)^2+(a_2 \sigma ^2)^2+(a_3 \sigma ^3)^2,\label{line3}
\ee
with three positive constants $a_1=\dpp a,\,a_2=\dpp b,\,a_3=\dpp c.$ In some cases, e.g. Bianchi I, these constants are not essential, i.e. they can be put to one by a change of coordinates. In the case of Bianchi V, only $c$ is essential. For Bianchi IX, all three constants are essential. 

When passing back to the 
Bianchi universes, the constants will become functions of time, the `scale factors', and all three will be essential in all nine cases.

The line elements (\ref{line3}) admit three Killing vectors $\xi _i,\,i=1,2,3,$
which are independent of the constants $a_i$. Indeed, already the three Maurer-Cartan forms are left unchanged by the three Killing vectors,
\bb
 {\cal L}_{\xi_i}\,\sigma^j=0,\qq i,j=1,2,3, \label{Maurer}
 \ee
 where $ {\cal L}_v$ is the Lie derivative with respect to a vector field $v$.

The three Killing vectors $\xi _i$ are listed in the second column and the non-vanishing structure constants ${C^k}_{ij}$ defined by $[\xi _i,\xi _j]=\dpp {C^k}_{ij}\,\xi _k$ are listed in the third column (up to antisymmetry in the lower indices $i,j$):
\bb 
\text{I}&
\pa_ x,\,\pa_ y,\,\pa_ z&{C^k}_{ij}=0\label{structure}\\
\text{II}&
\pa _x,\,\pa _y-z\pa _x,\,\pa _z&{C^1}_{23}=1 \nonumber\\
\text{III}&
\pa _x,\,\pa _y,\,x\pa x+\pa _z
&
{C^1}_{13}=1
\nonumber\\
\text{IV}&
\pa _x,\,\pa _y,\,(x+y)\,\pa _x+y\,\pa _y+\pa_z
& {C^1}_{13}={C^1}_{23}={C^2}_{23}=1\nonumber\\
\text{V}&
\pa _x,\,\pa _y,\,x\,\pa _x+y\,\pa _y+\pa_z
& {C^1}_{13}={C^2}_{23}=1
\nonumber\\
\text{VI}_h&
\pa _x,\,\pa _y,\,x\,\pa _x+hy\,\pa _y+\pa_z
& {C^1}_{13}=1,\,{C^2}_{23}=h\nonumber\\
\text{VII}_h&
\pa _x,\,\pa _y,\,x\,(hx+y)\,\pa _x+(-x+hy)\,\pa _y+\pa_z
& 
{C^1}_{23}={C^2}_{31}=1,{C^1}_{13}={C^2}_{23}=h\nonumber\\
\text{VIII}&
\cosh y\,\pa _x-\tanh x\,\sinh y\,\pa _y+\sinh y/\cosh x\,\pa z,\,\pa y,\,\
&\nonumber\\
&\hspace{-9mm}
 \sinh y\,\pa _x-\tanh x\,\cosh y\,\pa _y+\cosh y/\cosh x\,\pa z
 &
{C^1}_{23}={C^2}_{31}=1,\, {C^3}_{12}=-1\nonumber\\
\text{IX}&
\cos y\,\pa _x+\tan x\,\sin y\,\pa _y-\sin y/\cos x\,\pa z,\,\pa y,\,\
&\nonumber\\
&\hspace{-9mm}
 \sin y\,\pa _x-\tan x\,\cos y\,\pa _y+\cos y/\cos x\,\pa z
 &
{C^1}_{23}={C^2}_{31}={C^3}_{12}=1\nonumber
\ee
Note that, depending on the values of the constants $a_i$, there may be additional, independent Killing vectors, for example in the axial cases ($a=b$) or the maximally symmetric cases ($a=b=c$).

The three forms $\sigma ^i$ satisfy the Maurer-Cartan equations,
\bb
\de \sigma ^k={\textstyle\frac{1}{2}}\, {C^k}_{ij}\,\sigma ^i\wedge\sigma ^j,
\ee
with the same structure {\it constants} as defined by the commutators of Killing vectors and therefore the Ricci tensor with respect to the frame $\sigma ^i$ is constant, i.e. independent of the coordinates $x^i$. For convenience we use the orthonormal frame $e^i\dpp = a_i \sigma ^i$, no summation, with $g(e^i,e^j)= \delta ^{ij}$ and 
\bb
\de e ^k={\textstyle\frac{1}{2}}\, \hbox{${\hat C}^k$}_{ij}\,e ^i\wedge e ^j,&&
\hbox{${\hat C}^k$}_{ij}\dpp=\,\frac{a_k}{a_i\,a_j}\, {C^k}_{ij},\label{two}
\ee
 no summation in the second of equations (\ref{two}). In the orthonormal frame $e^i$, the Ricci tensor reads \cite{steph}:
 \bb
 R_{ij}=-{\textstyle\frac{1}{2}} \,\hbox{${\hat C}^k$}_{\ell i}\,\hbox{${\hat C}^\ell$}_{kj}
 -{\textstyle\frac{1}{2}} \,\hbox{${\hat C}^k$}_{\ell i}\,\hbox{${\hat C}^k$}_{\ell j}
 +{\textstyle\frac{1}{4}} \,\hbox{${\hat C}^i$}_{k\ell}\,\hbox{${\hat C}^j$}_{k\ell}
 -{\textstyle\frac{1}{2}} \,\hbox{${\hat C}^k$}_{k\ell }\lp\hbox{${\hat C}^i$}_{j\ell}+\hbox{${\hat C}^j$}_{i\ell}\rp,
 \ee
 summation over repeated indices irrespective of their vertical positions. 
 This explicit algebraic formula together with the list of structure constants (\ref{structure}) allows us to compute the Ricci tensors in the orthonormal frame $e^i$:
 \bb
\text{I}&0\,\text{diag}(1,1,1)&\nonumber\\[2mm]
\text{II}&{\textstyle\frac{1}{2}} \,\frac{a^2}{b^2c^2}\,  \text{diag}(1,-1,-1)&
\nonumber\\[2mm]
 \text{III}&
 -\,\frac{1}{a^2}\, \,\text{diag}(1,0,1)
 \nonumber\\[2mm]
 \text{IV}&
  - c^{-2}
 \begin{pmatrix}
2-{\textstyle\frac{1}{2}} r^2&r&0\\
r&2+{\textstyle\frac{1}{2}} r^2&0\\
0&0&2+{\textstyle\frac{1}{2}} r^2\end{pmatrix}, \ r\dpp =\,\frac{a}{b}\, 
&
 \nonumber\\[2mm]
\text{V}&-2\,\frac{1}{c^2}\, \text{diag}(1,1,1)&
\nonumber\\[2mm]
\text{VI}_h&-\,\frac{1}{c^2}\,\text{diag}(1+h,h(1+h),1+h^2)&
\nonumber\\[2mm]
\text{VII}_h&\hspace{5mm}
-{\textstyle\frac{1}{2}} c^{-2}
\begin{pmatrix}
4h^2-(r^2-1/r^2)&2h(r-1/r)&0\\
2h(r-1/r)&4h^2+(r^2-1/r^2)&0\\
0&0&4h^2+(r-1/r)^2\end{pmatrix}
\nonumber\\[2mm]
\text{VIII}&{\textstyle\frac{1}{2}} \,\frac{1}{a^2b^2c^2}\, \,\text{diag}(a^4-(b^2+c^2)^2,b^4-(a^2+c^2)^2,c^4-(a^2-b^2)^2)&
\nonumber\\[2mm]
\text{IX}&{\textstyle\frac{1}{2}} \,\frac{1}{a^2b^2c^2}\, \,\text{diag}(a^4-(b^2-c^2)^2,b^4-(a^2-c^2)^2,c^4-(a^2-b^2)^2)&
\eee
As in the orthonormal frame $e^i$ the metric tensor is equal to $\text{diag}(1,1,1)$ the Bianchi spaces I, V, VII$_h$ and IX with $a=b=c$ are Einstein and therefore maximally symmetric. They are close to maximally symmetric  if $b=a[1+\epsilon]$ and $c=a[1+\eta],\ |\epsilon|,|\eta|\ll 1$.

On the other hand, the Bianchi spaces II, III, IV, VI$_h$ and VIII do not contain any maximally symmetric space.
 
Finally there remains the question: Are there 4-dimensional Lie algebras admitting  representations as Killing vectors on 3-spaces that are not contained in the nine Bianchi spaces? Bianchi  answers the question to the negative by proving \cite{Bianchi} that
all 4-dimensional Lie algebras admitting  representations as Killing vectors on 3-spaces do have 3-dimensional Lie subalgebras. (In fact, any real or complex 4-dimensional Lie algebra has a 3-dimensional ideal.)

For completeness we must mention the Kantowski-Sachs metrics \cite{ks}:
\bb
\de s^2=a^2\, \de x ^2+b^2\,(\de \theta ^2+\sin^2 \theta \,\de\varphi ^2),\qq
\de s^2=a^2\, \de x ^2+b^2\,(\de \theta ^2+\sinh^2 \theta \,\de\varphi ^2).
\ee
They have 4-dimensional isometry groups and their corresponding universes have been analysed by Kantowski \& Sachs \cite{ks}.
These universes are usually not addressed as Bianchi universes, because their isometry groups do not contain 3-dimensional simply transitive sub-groups. More importantly for our criterium is the absence of Robertson-Walker universes in the Kantowski-Sachs families of universes.

Therefore minimal symmetry breaking of the cosmological principle allows for only three examples: the axial Bianchi I, V and IX universes.

Since
(axial and tri-axial) Bianchi~I universes have been studied extensively in the context of the Lema{\^i}tre-Hubble diagram \cite{stv}-\cite{fpu}, we remain here with
only two cases: the axial Bianchi V universes and the axial Bianchi IX universes. 

\section{Bianchi V}

In axial Bianchi V universes
 we encounter a well-known complication
 with the comobility of dust.
  
  Already in 1967 Farnsworth \cite{farn} has shown that axial Bianchi V universes do not support Einstein's equations with cosmological constant and comoving dust, unless they are already Friedman universes. In the wording of King \& Ellis \cite{KingEllis}, axial Bianchi V universes are `tilted' (unless Friedman). 
  
 Peculiar velocities of galaxies (modeled by dust) and a peculiar velocity of the observer (inducing a `kinematic dipole') are commonly considered {\it geographic} deviations from maximal symmetry. On the contrary, the flow of dust in axial Bianchi V universes is unavoidable and therefore a {\it geometric} deviation. 
      These universes were analyzed recently by Krishnan, Mondol and Sheikh-Jabbari \cite{sheik}.

 Farnsworth starts his proof by solving the Einstein equations with not necessarily comoving dust in axial Bianchi V universes. Then by inspecting the list of solutions, he finds that only the Friedman solutions allow for comoving dust.
 
 In subsection \ref{solutions} we give an alternative proof. Extending earlier results on vacuum solutions \cite{gal}, we
  solve the Einstein equations with  comoving dust in not necessarily axial Bianchi V universes. Then by inspecting the list of solutions, we find that only the Friedman solutions allow for axiality. 
 
 {\it En passant} we obtain an exact solution of the tri-axial Bianchi V universe with comoving dust discussed by Akarsu et al. \cite{ak}.

\subsection{Metric and isometries \label{isom}}
In order to have a simple approach to the isometries, we will write the Bianchi V metric using the Maurer-Cartan forms
\bb
\si_1=e^{-z}\,\de x,\qq\qq\qq \si_2=e^{-z}\,\de y,\qq\qq\qq \si_3=\de z,
\ee
so that we have
\bb\label{met5}
\de\tau^2=\de t^2-a(t)^2\,\si_1^2-b(t)^2\,\si_2^2-c(t)^2\,\si_3^2,\qq\qq
(x,y,z)\in\rr^3.
\ee 
The maximally symmetric case, $a=b=c$,  has  the isometry group $O(3,1)$ generated by the six Killing vectors,
\bb\barr{l}
\xi _1=\pt_x, \qq\qq \xi _2=\pt_y, \qq\qq \xi _3=x\pt_x+y\pt_y+\pt_z,\qq K_1=x\pt_y-y\pt_x,\\[5mm] K_2=\frac 12(e^{2z}-x^2+y^2)\pt_x-xy\pt_y-x\pt_z,\qq K_3=-xy\pt_x+\frac 12(e^{2z}+x^2-y^2)\pt_y-y\pt_z.\earr
\ee
The simplicity of the first four Killing vectors is a good reason for writing the Bianchi V metric in the coordinates $x,\,y,\,z$  rather than in the familiar polar coordinates $r ,\,\theta ,\,\varphi $.
We have already noted that the Maurer-Cartan forms are invariant under the first three Killing vectors, equation (\ref{Maurer}).
The remaining Killing vectors act on the Maurer-Cartan forms according to
\bb\barr{c}
{\cal L}_{K_1}\,\si_1=\si_2,\qq {\cal L}_{K_1}\,\si_2=-\si_1, \qq {\cal L}_{K_1}\,\si_3=0,\\ [5mm]
{\cal L}_{K_2}\,\si_1=y\,\si_2+e^{z}\,\si_3,\qq {\cal L}_{K_2}\,\si_2=-y\,\si_1,\qq {\cal L}_{K_2}\,\si_3=-e^{z}\,\si_1,\\ [5mm] {\cal L}_{K_3}\,\si_1=-x\,\si_2, \qq {\cal L}_{K_3}\,\si_2=x\,\si_1+e^{z}\,\si_3, \qq {\cal L}_{K_3}\,\si_3=-e^{z}\,\si_2.\earr
\ee 
Therefore if $a\not=b$ our metric (\ref{met5}) has three Killing vectors: $\xi _1,\,\xi _2,\,\xi _3$. If $a=b\not= c$ it has four Killing vectors:  $\xi _1,\,\xi _2,\,\xi _3$ and $K_1$, the axial case.

\subsection{Einstein equations with comoving dust \label{solutions}}
It is possible to obtain exact solutions of the Einstein equations with comoving dust for the general Bianchi V metric (\ref{met5}). To this end it is convenient to start from the following form of the metric 
\bb
\de\tau^2=\alf^2(\tg)\de\tg^2-a(\tg)^2\,\si_1^2-b(\tg)^2\,\si_2^2-c(\tg)^2\,\si_3^2,
\ee
so that we recover the cosmic time from $\de t=\alf(\tg)\de\tg$.

Until the end of this subsection we will use the prime to denote ordinary derivatives with respect to $\tg$ and the notations $H_a=a'/a,\ H_b=b'/b,\ H_c=c'/c$ and $H_{\alf}=\alf'/\alf$. Then the Einstein equations have a highly symmetric form
\bb\label{eqE}\barr{ll}
(a): &\qq\dst H'_a+H_a(H_a+H_b+H_c-H_{\alf})-2\frac{\alf^2}{c^2}=\Lambda\,\alf^2+4\pi G\rho\,\alf^2,\\[4mm]
(b): & \qq\dst H'_b+H_b(H_a+H_b+H_c-H_{\alf})-2\frac{\alf^2}{c^2}=\Lambda\,\alf^2+4\pi G\rho\,\alf^2,\\[4mm]
(c) &\qq\dst H'_c+H_c(H_a+H_b+H_c-H_{\alf})-2\frac{\alf^2}{c^2}=\Lambda\,\alf^2+4\pi G\rho\,\alf^2,\\[4mm]
(d) &\qq\dst H'_a+H'_b+H'_c+H_a^2+H_b^2+H_c^2-H_{\alf}(H_a+H_b+H_c)=\Lambda\,\alf^2-4\pi G\rho\,\alf^2,\\[5mm]
(e) &\qq\dst H_a+H_b=2H_c.\earr \label{off}
\ee
From the divergence of the energy-momentum tensor we get
\bb
\rho=\rho_0\frac{a_0\,b_0\,c_0}{abc}.\ee
 
The key to exact solutions is a good choice of $\alf$:  $\alf=abc$. Computing the differences (a)-(c) and (b)-(c) and integrating once we obtain
\bb
\frac{a'}{a}=\frac{c'}{c}+K, \qq\qq\qq \frac{b'}{b}=\frac{c'}{c}+L,\ee
while relation (e) implies $K+L=0$. Integrating both relations gives
\bb
a=\lambda\,e^{K\tg}\,c,\qq\qq b=\mu\,e^{-K\tg}\,c,\qq\qq \alf=\lambda\,\mu\,c^3,
\ee
and by a scaling of the coordinates $(x,y)$ we can set 
$\lambda=\mu=1$.

Then from relation (c) we have
\bb
\left(\frac{c'}{c}\right)'=
\Lambda\,c^6+2c^4+\frac R2\,c^3,\qq\qq\qq R\dpp=8\pi\,G\,\rho_0\,a_0b_0c_0.
\ee
Multiplying by $2\,c'/c$ and integrating leads to
\bb
\left(\frac{c'}{c}\right)^2=P(c), \qq\qq P(c)\dpp=\frac{\Lambda}{3}c^6+c^4+\frac R3\,c^3+E. \label{ode5}
\ee
Then relation (d) gives $E=K^2/3$
\bb
\de\tg^2=c^6\,\de\tg^2-c^2(e^{K\tg}\,\si_1^2+e^{-K\tg}\,\si_2^2+\si_3^2).
\ee
Using equation (\ref{ode5}), we obtain a differential equation for cosmic time $t$ as a function of the scale factor $c$,
\bb
\de t=c^3\,\de\tg=\frac{c^2}{\sqrt{P(c)}}\,\de c,
\ee
which can be integrated at the expense of genus 3 Abelian functions and inverted to obtain $c(t)$. Then, similarly, we have
\bb
\de\tg=\frac{\de c}{c\,\sqrt{P(c)}},
\ee
and an integration followed by an inversion will give $\tg(c)=\tg(c(t))$, leading to the final form of the metric
\bb
\de\tau^2=\de t^2-c^2(t)(e^{K\tg(c(t))}\,\si_1^2+e^{-K\tg(c(t))}\,\si_2^2+\si_3^2).
\ee
This general solution exhibits a very special feature: if we want axial symmetry, we must take $K=0$, therefore
\bb
\de\tau^2=\de t^2-c^2(t)\lp e^{-2z}\,\de x^2+e^{-2z}\,\de y^2+\de z^2\rp
\ee
and using the change of coordinates 
\bb
x=\frac rD\,\sin\tht\,\cos\varphi,\quad y=\frac rD\,\sin\tht\,\sin\varphi,\quad z=-\ln D,\quad D\dpp=\sqrt{1+r^2}-r\,\cos\tht ,
\ee
given in reference \cite{Barrow:1997vu} we end up with
\bb
\de\tau^2=\de t^2-c^2(t)\left(\frac{\de r^2}{1+r^2}+r^2(\de\tht^2+\sin^2\tht\,\de\varphi^2)\right)
\ee
and the Friedman equation
\bb
\left(\frac{1}{c}\, \frac{\de{c}}{\de t }\right)^2=\frac{\Lambda}{3}+\frac 1{c^2}+\frac{R}{3c^3}\,.
\ee

In conclusion axial Bianchi V universes are incompatible with Einstein's equation for comoving dust.

\subsection{Two scale factors with comoving dust}

Axial Bianchi V universes are obtained from generic Bianchi V universes by reducing the three independent scale factors, $a,\,b,$ and $c$ to two by setting $b=a$. This choice increases the number of Killing vectors from three to four. But at the same time it is incompatible with comoving dust in Einstein's equation. 

Recently Akarsu et al. \cite{ak} have introduced a different reduction to two independent scale factors: $c^2=ab$ with $a\not=b$. This reduction is equivalent to the off-diagonal  Einstein equation (\ref{off}\,e). 
Therefore the model has comoving dust and our exact solution above applies.
 
However, as we have shown in subsection 
\ref{isom}, this model has only three Killing vectors and does not qualify as minimal symmetry breaking of the cosmological principle.

\section{Axial Bianchi~IX\label{b9}}

Bianchi IX universes have compact, simply connected, homogeneous 3-spaces. The isometry group, $SO(3)$, is nonAbelian and compact, on the  diametrical opposite of the Abelian isometry group of Bianchi I. Much is known about the dynamics of Bianchi IX universes.

Wald (1983): {\it The far future} of all initially expanding Bianchi I - VIII universes satisfying Einstein's equations with positive cosmological constant approaches the de Sitter universe exponentially fast. This only remains true for Bianchi IX universes if the cosmological constant is sufficiently large with respect to spacial curvature. ``A positive cosmological constant provides an effective means of isotropizing homogeneous universes.''

In {\it the far past}, Bianchi IX universes have complicated dynamics, oscillatory singularities, and chaos \cite{misn,bkl1,bkl2}. Intriguingly the axial Bianchi IX universes are better behaved than the tri-axial ones.
Dechant, Lasenby \& Hobson \cite{dlh1,dlh2}
add a scalar field to axial Bianchi IX universes in order to soften the initial singularity.
Let us stress that we will modestly remain in  {\it our recent past}, $z<6$.

Giani, Piattella \& Kamenshchik \cite{gpk} have used Bianchi IX universes  to model the gravitational collapse of matter inhomogeneities. They
 map the Bianchi IX field equations into those for a spherical Friedman universe filled with two scalar fields, only one scalar field in the axial case. 
 We will meet the Klein-Gordon equation of this scalar field when we linearize the anisotropy, equation (\ref{eta9}). 
 
 Our theoretical motivations for considering axial Bianchi IX universes is the minimal symmetry breaking of the cosmological principle together with Farnsworth's theorem from 1967. On the observational side we note that a  Friedman universe with small positive curvature is still compatible with data \cite{uke, div, lhtw}
 and we think it is worthwhile to consider a model featuring both small positive curvature and small anisotropies.
 
\subsection{Metric}
The Bianchi IX universe can be defined \cite{Bianchi,ter} by:
\begin{align}
\de \tau^2= \de t^2&-{\textstyle\frac{1}{4}} a(t)^2(\cos z\, \de x-\sin z \cos x\, \de y)^2
\nonumber\\[2mm]&
-{\textstyle\frac{1}{4}}b(t)^2(\sin z \,\de x+\cos z \cos x \,\de y)^2
-{\textstyle\frac{1}{4}}c(t)^2(\de z+\sin x \,\de y)^2,
\end{align}
with the Euler angles $x\in ({\textstyle\frac{3}{2}}\pi  ,{\textstyle\frac{3}{2}}\pi+2\pi  )$, $y\in (0,2\pi )$ and $z\in (0,2\pi )$. It has three Killing vectors:
\bb
\cos y\,\pa_x+\tan x\,\sin y \,\pa_y \,-\,\frac{\sin y}{\cos x}\ \pa_z,\ \pa _y,\,
\sin y\,\pa_x-\tan x\,\cos y \,\pa_y \,+\,\frac{\cos y}{\cos x}\, \pa_z.
\ee
The axial Bianchi IX universe is obtained by setting $b=a$:
\begin{align}
\de \tau^2= \de t^2&-{\textstyle\frac{1}{4}} a^2 \de x^2-{\textstyle\frac{1}{4}}(a^2\cos^2x+c^2 \sin^2x)\,\de y^2
-{\textstyle\frac{1}{4}}c^2\,\de z^2-{\textstyle\frac{1}{2}} c^2\sin x\,\de y\,\de z\,.
\end{align}
It has a fourth Killing vector, $\pa_z$. 

Setting $c=a$ we arrive at the spherical Robertson-Walker universe expressed in Euler angles:
\begin{align}
\de \tau^2= \de t^2&-{\textstyle\frac{1}{4}} a^2 (\de x^2+\de y^2
+\de z^2+2\sin x\,\de y\,\de z\,)
\end{align}
 with its six Killing vectors generating the mentioned isometry group $O(4)$. 

The Euler angles $x,y,z$ are related to the familiar polar coordinates by \cite{Ki}:
\bb
\chi&=&\arccos\lp\cos\frac{x-{\textstyle\frac{3}{2}}\pi }{2}\, \sin\frac{y-z}{2}\rp , \label{chi} \\[2mm]
\theta &=&\arctan\lp\tan\frac{x-{\textstyle\frac{3}{2}}\pi }{2}\,/\cos\frac{y-z}{2}\rp ,\label{theta}\\[2mm]
\varphi &=&\,\frac{y+z}{2}\,, \label{phi}
\ee
and
\bb
x&=&2\,\arccos\sqrt{1-\sin^2\chi\,\sin^2\theta }+{\textstyle\frac{3}{2}}\pi ,\label{xfrompolar}\\[2mm]
y&=&\varphi + \arccos\,\frac{\sin\chi\,\cos\theta }{\sqrt{1-\sin^2\chi\,\sin^2\theta }}\,,\label{yfrompolar}\\[2mm] 
z&=&\varphi - \arccos\,\frac{\sin\chi\,\cos\theta }{\sqrt{1-\sin^2\chi\,\sin^2\theta }}\,.\label{zfrompolar}
\ee
 The metric tensor reads,
\bb
g_{\mu \nu }=
\begin{pmatrix}
1&0&0&0\\
0&-{\textstyle\frac{1}{4}} a^2&0&0\\
0&0&-{\textstyle\frac{1}{4}}(a^2 \cos^2x+c^2\sin ^2x)&-{\textstyle\frac{1}{4}}c^2\sin x\\
0&0&-{\textstyle\frac{1}{4}}c^2\sin x&-{\textstyle\frac{1}{4}}c^2
\end{pmatrix}\,.\label{metric}
\ee
Note the relations, $g_{yy}=\cos^2x\,g_{xx}+\sin^2x\,g_{zz}$ and $g_{yz}=\sin x\,g_{zz}$.
The inverse metric tensor is:
\bb
g^{\mu \nu }=
\begin{pmatrix}
1&0&0&0\\
0&-4a^{-2}&0&0\\
0&0&-4a^{-2} \cos^{-2}x&4a^{-2} \cos^{-2}x\,\sin x\\
0&0&4a^{-2} \cos^{-2}x\,\sin x&-4(a^{-2} \cos^{-2}x\,\sin^2x+c^{-2})
\end{pmatrix}\,.
\ee

\subsection{Geodesics \label{aa9}}

Now we are ready to prove that our coordinates are comoving with respect to test masses.

With the four Killing vectors, Emmy Noether's theorem yields four conserved quantities on any geodesic $x^\mu (q)$. We can compute them conveniently by defining a Hamiltonian $\hh(x,p)\dpp={\textstyle\frac{1}{2}} g^{\mu \nu }(x)\,p_\mu p_\nu$ and by using Hamilton's equations with $\dot\ \dpp= {\de}/{\de q}$,
\bb
\dot x^\mu &=&\ \ \frac{\pa \hh}{\pa p_\mu }\,=\,p^\mu ,\label{lin}\\[2mm]
\dot p_\mu&=&-\,\frac{\pa \hh}{\pa x^\mu }\,.\label{lin2}
\ee
We interpret equation (\ref{lin})  as definition of the 4-momentum $p_\mu $; then equation (\ref{lin2}) is nothing but the geodesic equation. For the conserved quantity generated by a Killing vector $\xi =\xi ^\mu \pa_\mu $ we may take $\xi ^\mu p_\mu$. 
The four Killing vectors,
\begin{align}
&\sin y\,\pa_x-\,\frac{\sin x}{\cos x}\,\cos y \,\pa_y \,+\,\frac{\cos y}{\cos x}\, \pa_z,
\nonumber\\[2mm]
&\cos y\,\pa_x+\,\frac{\sin x}{\cos x}\,\sin y \,\pa_y \,-\,\frac{\sin y}{\cos x}\, \pa_z,\qq\pa_y\qq\text{and}\qq\pa _z,
\end{align}
yield the four conserved quantities  (up to a factor $-{\textstyle\frac{1}{4}}$, that we drop):
\bb
A&\dpp=&a^2 \sin y\,\dot x+(c^2-a^2)\,{\sin x}\,{\cos x}\,\cos y\,\dot y+c^2\,{\cos x}\,{\cos y}\,\dot z\,,\label{conserved1}\\
\bar A &\dpp=& a^2 \cos y\,\dot x-(c^2-a^2)\,{\sin x}\,{\cos x}\,\sin y\,\dot y-c^2{\cos x}\,{\sin y}\,\dot z\,,\\
B&\dpp=&\hspace{2.7cm}(a^2\cos^2x+c^2\sin^2x)\,\dot y+c^2\sin x\,\dot z,
\label{conserved3}\\
C&\dpp=& \hspace{5.2cm}c^2\sin x \,\dot y+c^2\dot z\,,\label{conserved4}
\ee
or equivalently
\bb
\dot x &=&\pm\,\frac{1 }{a^2}\lb
A^2+\bar A^2-\lp \frac{B\sin x-C}{\cos x} \rp^2\rb^{1/2}\,,
\label{xdot}\\[2mm]
\dot y&=&\,\frac{B-C\sin x}{a^2\cos^2 x}\, ,\label{ydot}\\[2mm]
\dot z&=&\,\frac{C}{c^2}\, -\sin x\,\frac{B-C\sin x}{a^2\cos^2 x}\, ,
\label{zdot}\\[2mm]
 A\cos y-\bar A \sin y &=& -\,\frac{B\,\sin x-C}{\cos x}\, \label{star}. 
\ee
Any dust test-particle with vanishing initial 3-velocity, $A=B=C=0$,  will remain at rest with respect to the coordinates $(x,y,z)$.

For a massless test particle, we have $\dot x^\mu g_{\mu \nu }\dot x^\nu=0$ or equivalently:
\bb
  \dot t^2=\,\frac{A^2+\bar A^2+B^2-C^2}{4\,a^2}\,+\,\frac{C^2}{4\,c^2}\, 
  \label{tdot} .
 \ee
 
 \subsection{Redshift}
 We are now ready to compute the redshift in axial Bianchi IX universes. Since every Bianchi space is homogeneous (in the sense that it forms a single orbit under the group action), we may place the emitter at the north pole, $\chi_e=0$. This choice considerably simplifies the computations, but we must pay due attention to the coordinate singularity at the poles. Equations (\ref{xfrompolar})-(\ref{zfrompolar}) fix $x_e={\textstyle\frac{3}{2}} \pi $ and $z_e=y_e\pm\pi $.
 
 At cosmic time $t_e$ our emitter at the north pole radiates a photon. As before we call $q$ the affine parameter of the light-like geodesic and $\dot\ \dpp= {\de}/{\de q}$. To have a unique solution of the geodesic equation in the interval $[q_e,q_0]$ we choose initial conditions at $q_e$:
\bb
\begin{tabular}{llll}
 $t(q_e)=t_e$,&  $x(q_e)=x_e={\textstyle\frac{3}{2}} \pi$,&
   $y(q_e)=y_e$,&$z(q_e)=z_e=y_e+\pi$,
 \\[2mm]
 $\dot t(q_e)=\dot t_e>0$,& $\dot x(q_e)=R>0$,&
$\dot y(q_e)=S$,& 
 $\dot z(q_e)=\dot z_e$.
 \end{tabular}\label{initial}
\ee
They determine the conserved quantities by equations (\ref{conserved1})-(\ref{conserved4}):
\bb 
\,\frac{A}{\sin y_e}\, =\,\frac{\bar A}{\cos y_e}\,=a_e^2\,R,\qq
A^2+\bar A^2=a_e^4\,R^2,\\[2mm]
B=-C=c_e^2(S-\dot z_e),
\ee
and evaluating equation (\ref{zdot}) at emission we obtain the initial condition
\bb
\dot z_e=\lb 1-2\,\frac{a_e^2}{c_e^2}\rb S,
\ee 
where we have used the short-hand $a_e\dpp=a(t_e)$, $c_e\dpp=c(t_e)$.
Other abbreviations will be useful:  
\bb 
s\dpp=\,\frac{2S}{R}\,,\
W(t)\dpp=2\lb \frac{1}{a^2}\,+\,\frac{s^2}{c^2}\rb^{-1/2},\
f(x)\dpp=\,\frac{1+\sin x}{\cos x}\,,\
V(x)\dpp = \lb 1-s^2f(x)^2\rb^{-1/2};\label{4defs}
\ee
as well as the derivatives,
\bb
f_x(x)\dpp=\,\frac{\de}{\de x}\,f(x) =\,\frac{1+\sin x}{\cos^2 x}\,,\qq
\,\frac{\de}{\de x}\,f_x(x)=f(x)\,f_x(x),\ \dots\, ;
\ee
the relations,
\bb
2\,f_x=1+f^2\,,\qq f(x)=\tan\frac{x-{\textstyle\frac{3}{2}}\pi }{2}\,,\qq f_x(x)=
\,\frac{1}{1-\sin x}\,; \label{3rels}
\ee
the integrals,
\bb
\int V=\,\frac{2}{\sqrt{1+s^2}}\, \arctan(\sqrt{1+s^2}\,f\,V),\qq
\int(1+f^2)\,V^3=2\,f\,V;\label{2ints}
\ee
and the limits at the north pole,
\bb
\lim_{x\rightarrow x_e}f(x)=0,\qq \lim_{x\rightarrow x_e}f_x(x)={\textstyle\frac{1}{2}},\ \dots\, . \label{limits}
\ee
In these notations, equations (\ref{tdot}), (\ref{xdot})-(\ref{star}) become
\bb \dot t&=&a_e^2\,R\,W^{-1},\qq
\dot x\ =\ \,\frac{a_e^2}{a^2}\, R\,V^{-1},\qq
\label{tdotxdot}\\[2mm]
\dot y&=&\frac{a_e^2}{a^2}\, R\,s\,f_x,\qq
\dot z\ =\ -\dot y+a_e^2\,R\,s\lb\frac{1}{a^2}\,- \,\frac{1}{c^2}\rb,
\label{ydotzdot}\\[2mm]
&&\sin(y-y_e)\ =\ s\,f(x). \label{sstar}
\ee
Let us compute the initial polar angles (\ref{theta}) and (\ref{phi}). For $\theta $ we use  equation (\ref{sstar}) and write
\bb
\tan\theta =\,\frac{f}{\cos\frac{y-z}{2}}\, =\,\frac{1}{s}\,\frac{\sin(y-y_e)}{\cos\frac{y-z}{2}}\,  =\,\frac{1}{s}\,\frac{\cos(y-y_e-{\textstyle\frac{1}{2}} \pi )}{\cos\frac{y-z}{2}}\,.  \label{theta2}
\ee
 In this last form we can evaluate $\tan \theta $ safely at the north pole and find $\tan\theta _e=1/s={\textstyle\frac{1}{2}} \dot x_e/\dot y_e$. For $\varphi $ we obtain $\varphi _e=(y_e+z_e)/2=y_e+\pi /2$.

Note that deriving equation (\ref{sstar}) with respect to the affine parameter $q$ and using the equation for $\dot x$ in (\ref{tdotxdot}) reproduces the equation for $\dot y$ in (\ref{ydotzdot}), which we may therefore drop. At the same time we will trade the equation for $\dot z$ in favor of the equation for $\dot\varphi = (\dot y+\dot z)/2$.

Next we eliminate the affine parameter $q$ in favor of cosmic time $t$ and remain with only three equations,
\bb
\,\frac{\de x}{\de t}\, =\,\frac{1}{a^2}\,\frac{W}{V}\,,\qq\qq
 \,\frac{\de \varphi }{\de t}\, ={\textstyle\frac{1}{2}} \,s\lb\frac{1}{a^2}\,- \,\frac{1}{c^2}\rb W,\qq\qq
 \sin(y-y_e)\ =\ s\,f(x).\label{geodesic}
\ee
 We suppose that the photon arrives today, $t_0=t(q_0)$, in our telescope situated at $\vec x_0\dpp=\vec x(q_0)$ and integrate the first two of these three equations by separation of variables,
 \bb
 \int_{t_e}^{t_0}\,\frac{W}{a^2}\, =\int_{x_e}^{x_0}V
 =\,\frac{2}{\sqrt{1+s^2}}\, \arctan(\sqrt{1+s^2}\,f_0\,V_0),\label{firstin}
\\[2mm]
 \,\frac{s}{2}  \int_{t_e}^{t_0}\lb\frac{1}{a^2}\,- \,\frac{1}{c^2}\rb W
 =\varphi _0-\varphi _e=\varphi _0-y_e-{\textstyle\frac{1}{2}} \pi .\label{secondin}
 \ee
 
Consider now a second photon emitted at the north pole an atomic period $T$ later in cosmic time than the first photon and arriving in our comoving telescope still at $\vec x_0$ at cosmic time $t_0+T_D$. ($\cdot_D$ stands for Doppler.) We suppose that the atomic periods, $T$ and $T_D$, are much smaller than the time of flight $t_0-t_e$ and the geodesics of the two photons, $x^\mu (q)$ and $\tilde x^\mu (q)$, $q\in[q_e,q_0]$ are infinitesimally close. For the second geodesic, the initial conditions at $q_e$ are:
\bb
\begin{tabular}{llll}
 $\tilde t(q_e)=t_e+T$,&  $\tilde x(q_e)=x_e={\textstyle\frac{3}{2}} \pi$,&
   $\tilde y(q_e)=\tilde y_e$,&$\tilde z(q_e)=\tilde y_e+\pi$,
 \\[2mm]
 $\dot {\tilde t}(q_e)>0$,& $\dot {\tilde x}(q_e)=\tilde R>0$,&
$\dot{\tilde y}(q_e)=\tilde S$,& 
 $\dot {\tilde z}(q_e)=\lb 1-2\,\frac{\tilde a_e^2}{\tilde c_e^2}\rb \tilde S$.
 \end{tabular}
\ee
We want to adjust the initial `velocity' (or direction)
 of the second photon,
 \bb
 \tilde y_e=\dpp y_e[1+\delta  ],\qq\qq
  \tilde s\dpp=\,\frac{2\tilde S}{\tilde R}\,= \dpp s[1+\epsilon ],
 \ee
  such that its final comoving position coincides with that of the first photon, $\, \tilde {\vec x}_0= \vec x_0$. The adjustment being infinitesimal, we may keep only linear terms in $T,\,T_D,\,\delta  $ and $\epsilon $. In this approximation, that we also indicate by $\sim$,  equation (\ref{sstar}) yields
\bb
\delta  y_e\sim-\epsilon \,\frac{sf_0}{\sqrt{1-s^2f_0^2}}\,=-\epsilon sf_0V_0. 
\ee
With
  \bb
 W_\epsilon\dpp=2\lb \frac{1}{a^2}\,+\,\frac{\tilde s^2}{c^2}\rb^{-1/2}
 \sim W\lb1-\,\frac{\epsilon s^2}{4c^2}\,W^2\rb,\qq
  V_\epsilon\dpp = \lb 1-\tilde s^2f^2\rb^{-1/2}\sim V\lb1+\epsilon s^2f^2V^2\rb,
  \label{Weps}
  \ee 
the two equations (\ref{firstin}),(\ref{secondin}) yield,
\bb
\,\frac{1}{a_0^2}\,W_0\,T_D- \,\frac{1}{a_e^2}\,W_e\,T&\sim&
 \epsilon s^2\lp {\textstyle\frac{1}{4}} \int_{t_e}^{t_0}\,\frac{W^3}{a^2c^2}\, +\int_{x_e}^{x_0}f^2V^3\rp,\label{infx}\\[2mm]
\hspace{-7mm}
\lb\frac{1}{a_0^2}\,-\,\frac{1}{c_0^2}\rb W_0\,T_D- \lb\frac{1}{a_e^2}\,-\,\frac{1}{c_e^2}\rb W_e\,T&\sim&
 \epsilon   \int_{t_e}^{t_0}\lb\frac{1}{a^2}\,-\,\frac{1}{c^2}\rb \lp
{\frac{s^2}{4}}\,\frac{W^3}{c^2}\, - W\rp-2\delta \,\frac{y_e}{s}
\nonumber\\[2mm]&\sim&
- {\textstyle\frac{1}{4}} \epsilon   \int_{t_e}^{t_0}\lb\frac{1}{a^2}\,-\,\frac{1}{c^2}\rb\frac{W^3}{a^2}\, +2 \epsilon f_0V_0\label{inffi} .
\ee
Now we take $(1+s^2)$ times equation (\ref{infx}) and subtract $s^2$ times 
equation (\ref{inffi}). We obtain,
\bb
4\lp\frac{T_D}{W_0}\,- \,\frac{T}{W_e}\rp&\sim&
 \epsilon s^2\lp {\textstyle\frac{1}{4}}\int_{t_e}^{t_0}\lb\frac{1}{a^2}\,+\,\frac{s^2}{c^2}\rb\frac{W^3}{a^2}\, +(1+s^2)\int_{x_e}^{x_0}f^2V^3-2f_0V_0\rp
\nonumber\\[2mm]&=&
 \epsilon s^2\lp \int_{t_e}^{t_0}\frac{W}{a^2}\, +(1+s^2)\int_{x_e}^{x_0}f^2V^3-2f_0V_0\rp
\nonumber\\[2mm]&=&
 \epsilon s^2\lp \int_{x_e}^{x_0}V +(1+s^2)\int_{x_e}^{x_0}f^2V^3-2f_0V_0\rp
\nonumber\\[2mm]&=&
 \epsilon s^2\lp \int_{x_e}^{x_0}(1+f^2)V^3 -2f_0V_0\rp=
 \epsilon s^2\lp 2fV|_{x_e}^{x_0} -2f_0V_0\rp=0,
\ee
where we have used successively the definition of $W$ (the second of the four definitions (\ref{4defs})), equation (\ref{firstin}), the definition of $V$ (the fourth of the definitions (\ref{4defs})), the second of the two integrals (\ref{2ints}) and the first of the limits (\ref{limits}).

If we choose $T=0$, we learn that $ \epsilon =\delta  =0$, which means absence of weak lensing in axial Bianchi IX universes. Very strong lensing of course exists (as in spherical Friedman universes) for an observer at the south pole, $\chi_0 = \pi,\ x_0={\textstyle\frac{3}{2}} \pi +2\pi $, if she is not below the horizon.
For non-vanishing $T$, we have the redshift 
\bb 
 \ul z\dpp=\,\frac{T_D-T}{T}\,\sim  \,\frac{W_0}{W_e}\,-1.\label{redshift}
 \ee
 This formula agrees with the general definition of redshift, that you find in Pierre Fleury's  beautiful thesis \cite{thesis} as equation (1.71).
  
 To avoid confusion with the third Euler angle, we underline  the redshift $\ul z$ temporarily.

\subsection{Apparent luminosity}

The trajectory of our photon is described by the lightlike geodesic $\vec x(t)$ with initial conditions (\ref{initial}) and given by equations (\ref{geodesic}-\ref{secondin}). 
We obtained the redshift from the infinitesimal variation of the initial conditions in (\ref{initial}):
\bb
t_e\rightarrow t_e+T,\qq y_e\rightarrow y_e(1+\delta),\qq s\rightarrow s(1+\epsilon).
\ee
Requiring that the two geodesics end at the same point in space $\vec x_0={\tilde{\vec x}}_0$ we computed the proper time difference at reception $T_D=\tilde t_0-t_0$.
For the apparent luminosity we need two more infinitesimally close lightlike geodesics 
\bb 
\vec x_\delta(t)=\vec x(t)+\vec\delta (t)\qq \text{and} \qq \vec x_\epsilon(t)=\vec x(t)+\vec\epsilon (t) 
\ee
 with varied spatial initial conditions $y_e\rightarrow y_e(1+\delta)$ and $s\rightarrow s(1+\epsilon)$ respectively. All three geodesics are emitted simultaneously at $t_e$ from the point $\vec x_e$ . The two 3-vectors $\vec\delta(t) $ and $\vec\epsilon(t)$, which should not be confused with the infinitesimal numbers $\delta $ and $\epsilon$, are interpreteted as tangent vectors at $\vec x(t)$ just as the velocity $\vec x'(t)$. The prime denotes the derivative with respect to cosmic time $t$. As a function of time, $\vec\delta(t) $ and $\vec\epsilon(t)$ define a family of infinitesimal triangles. They are  perpendicular to the trajectory $\vec x(t)$. The triangles represent the beam of photons emitted by the supernova at $\vec x(t_e)$.  The photons illuminate a triangular photo-plate now and here  $\vec x(t_0)$, see figure 1. The plate is placed perpendicular to the beam and its area is  $\de S(t_0)=\dpp \de S_0$ . Our task is to compute the area $\de S(t)$ of the infinitesimal triangles and their solid angle $\de \Omega _e$ at emission. Then the apparent luminosity is:
 \bb
 \ell=L\,\frac{\de \Omega _e}{\de S_0}\, (\ul z+1)^{-2},
 \ee
 with the absolute luminosity $L$ of the supernova.\\
 \vspace{4mm}
 \begin{center}
\begin{tabular}{c}
\xy
(0,70)*{}="N";
(50,0)*{}="E";
(41.8,4)*{}="Ed";
(60,6)*{}="Ee";
(46.6,17.1)*{}="Px"; 
(48,17.6)*{}="P"; 
(48.5,15.5)*{}="Pee"; 
(45.6,17.9)*{}="Q";
(46.0,16.1)*{}="Qd"; 
"N"; "E" **\crv{(40,65)};
"N"; "Ed" **\crv{(40,55)};
"N"; "Ee" **\crv{(40,70)};
(47.2,15)*{}="T";
(40,21)*{}; 
(40,21)*{};="Td"; 
(55.7,18)*{}; 
(55.7,18)*{};="Tee"; 
(56.7,15)*{}; 
(56.7,15)*{};="Te";
{\ar "T"; "Td"};
{\ar "T"; "Tee"};
%"T"; "Tee"*{} **\dir{-};
"Td"; (45,20)*{} **\dir{-}; 
(47,19.6)*{}; "Tee"*{} **\dir{-}; 
"P"; "Px"*{} **\dir{-};
"P"; "Pee"*{} **\dir{-}; 
"Q"; "Px"*{} **\dir{-};
"Q"; "Qd"*{} **\dir{-}; 
(-3,71)*{N};
(41,-0.2)*{\vec x_\delta (t)};
(51,-3.7)*{\vec x(t)};
(63,2)*{\vec x_\epsilon(t)};
(43.5,13.9)*{\vec\delta };
(52.5,12.6)*{\vec\epsilon };
(8,67.5)*{.};
(7.7,67.2)*{}; (-5,62)*{} **\crv{(3,62)};
(-10,62)*{\de\Omega _e };
(28,69)*{\de S(t_e+\de t) };
(15.3,65.2); (16.,67.1)*{} **\dir{-}; 
(15.3,65.2); (10,65.3)*{} **\dir{-}; 
(10,65.3); (12.6,66.)*{} **\dir{-}; 
(14.,66.4);(16.,67.1) *{} **\dir{-}; 
(49.9,17.5)*{.};
(49.9,17.9)*{}; (68,25)*{} **\crv{(50,30)};
(75,24.7)*{\de S(t_0) };
(0,-15)*{};
\endxy
\end{tabular}\linebreak\nopagebreak
{Figure 1: Three infinitesimally close geodesics and two perpendicular triangles}
\end{center}
  \vspace{4mm}
  
Explicitely the three 3-vectors are:
\bb
\vec x'=\,\frac{sW}{a^2}\, 
\begin{pmatrix}
(sV)^{-1}\\
f_x\\
-f_x+\lb1-\,\frac{a^2}{c^2}\rb
\end{pmatrix},\qq
\vec\delta =\delta \,y_e
\begin{pmatrix}
0\\
1\\
1
\end{pmatrix},\qq\label{x'delta}\\[2mm]
\vec\epsilon=\epsilon\,\frac{s^2}{1+s^2}\la f
 \begin{pmatrix}
-2\\
(sV)^{-1}\\
-(sV)^{-1}
\end{pmatrix}+sI_3
\begin{pmatrix}
(sV)^{-1}\\
f_x\\
-f_x+\,\frac{(1+s^2)}{s^2}\, 
\end{pmatrix}\ra,
\label{epsilon}
\ee
 with
 \bb
 I_3(t)\dpp={\textstyle\frac{1}{4}} \int_{t _e}^t\frac{W^3}{a^2}\lb\frac{1}{a^2}-\,\frac{1}{c^2}\rb.\label{I3}
 \ee  
 The derivation of equations (\ref{x'delta}) is straight-forward. For equation (\ref{epsilon}) it is convenient to start with
 equation (\ref{firstin}) in the form
 \bb
 \int_{t_e}^{t}\,\frac{W}{a^2}\, =\int_{x_e}^{x}V  =\,\frac{2}{\sqrt{1+s^2}}\, \arctan(\sqrt{1+s^2}\,f\,V)=\dpp G(x,s),
 \ee
and to replace $s\rightarrow s(1+\epsilon)$ and  $x\rightarrow x+ \epsilon_x$,
\bb
 \int_{t_e}^{t}\,\frac{W_\epsilon}{a^2}\, = G(x+\epsilon_x,s+s\epsilon),
 \ee
with $W_\epsilon$ given by the first of equations (\ref{Weps}):
 \bb
 W_\epsilon\sim W\lb1-\,\frac{\epsilon s^2}{4c^2}\,W^2\rb.  \ee 
For the linearization we also need,
\bb
\pa _x\,G=V,\qq \text{and} \qq
\pa _s\,G=\,\frac{s}{1+s^2}\,(2fV-G), 
 \ee
and with the algebraic identity,
\bb
{\textstyle\frac{1}{4}} \,\frac{W^2}{c^2}\,-\,\frac{1}{1+s^2}\, =\,-\,\frac{1}{1+s^2}\, {\textstyle\frac{1}{4}}  W^2\lb\frac{1}{a^2}\,-\,\frac{1}{c^2}\rb,
\ee  
we obtain the $x$-component of $\vec\epsilon$ in equation (\ref{epsilon}).
The $y$- and $z$-components then are computed by linearizing the third of equations (\ref{geodesic}),
$ \sin(y-y_e)=sf(x)$, and equation (\ref{secondin}).
 
  Let us denote by $\vec\delta \cdot \vec\epsilon$ the negative of the scalar product of 3-vectors defined by the spatial part of the metric tensor $g_{\mu \nu }$ in equation (\ref{metric}). In particular $\vec x'\cdot \vec x'=1$, since we have set the speed of light to one. For the other five scalar products we find:  
  \bb
  \begin{split}
 \vec\delta \cdot\vec\delta\,&={\textstyle\frac{1}{4 }} \,\delta ^2\,y_e^2\, \cos^2x\,[a^2+ c^2f^2],\\[2mm]
  \vec\epsilon\cdot\vec\epsilon\,&=
      \,\frac{\epsilon^2s^2}{(1+s^2)^2}\Big\{  s^2a^2f^2+{\textstyle\frac{1}{4}} \lb a^2f^2+c^2\rb\,\frac{\cos^2x}{V^2}\,\\[2mm]& \hspace{40mm}
  -2I_3 \,\frac{a^2c^2}{W^2}\,\frac{\cos x}{V}\,
+I_3^2\,\frac{a^2c^2}{W^2}\Big\},\\[2mm]
  \vec\delta \cdot\vec\epsilon\,&=\,\frac{\delta\, \epsilon\, y_es}{(1+s^2)}\, \Big\{
  {\textstyle\frac{1}{4}}(a^2-c^2)\,\frac{\cos x}{V}\, +I_3 \,\frac{a^2c^2}{W^2} \Big\}(1+\sin x),\\[2mm]
 \vec\delta \cdot\vec x'&=0,\hspace{11mm}
  \vec\epsilon\cdot\vec x'=0.
 \end{split}
  \ee
  The infinitesimal triangles defined by $\vec\delta(t)$ and  $\vec\epsilon(t)$ are perpendicular to the ray $\vec x'(t)$ and their areas as a function of time $\de S(t)$, see figure 1, are given by
 \bb
 \de S&=&{\textstyle\frac{1}{2}} \lp(\vec \delta \cdot \vec \delta)  (\vec\epsilon\cdot\vec\epsilon)-
(\vec\delta \cdot\vec\epsilon)^2\rp^{1/2}\nonumber\\[2mm]
&=&{\textstyle\frac{1}{2}} \delta \epsilon\, y_e \,\frac{s}{1+s^2}\,\frac{a^2c}{W}\,(1+\sin x)\,\Big| 1-{\textstyle\frac{1}{2}}\,\frac{I_3}{fV} \Big|.   
\ee
Of course at emission time $t_e$ the area vanishes and to compute the solid angle it sustains at the north pole $\vec x(t_e)$ we evaluate $\de S(t_e+\de t)$ to leading order in $\de t$. From the first of equations (\ref{geodesic}) we have $\de x\sim W_e/a_e^2\,\de t$ 
and:
\bb
1+\sin x(t_e+\de t)\sim{\textstyle\frac{1}{2}} \,\frac{W_e^2}{a_e^4}\, \de t^2,\qq\qq
f(x(t_e+\de t))\sim{\textstyle\frac{1}{2}} \,\frac{W_e}{a_e^2}\, \de t.
\ee
 We linearize the integral $I_3$ defined in equation (\ref{I3}) with respect to $\de t$,
\bb
I_3(t_e+\de t)\sim{\textstyle\frac{1}{4}} \,\frac{W_e^3}{a_e^2}\lb\frac{1}{a_e^2}\,-\,\frac{1}{c_e^2}\, \rb\de t.
\ee  
 Therefore
\bb
1-{\textstyle\frac{1}{2}}\,\frac{I_3}{fV}\,=
{\textstyle\frac{1}{4}} \,\frac{1+s^2}{c_e^2}\, W_e^2\,+O(\de t)
 \qq
\text{and}\qq
\de S(t_e+\de t)\sim{\textstyle\frac{1}{16}} 
\delta \epsilon\,y_e s\,\frac{W^3_e}{a_e^2c_e}\, \de t^2.
\ee
Finally the solid angle is
\bb
\de\Omega _e=\,\frac{\de S(t_e+\de t)}{4\pi\, \de t^2}\,
 =\,\frac{\delta \epsilon}{64\pi }\,y_e s\,\frac{W^3_e}{a_e^2c_e}\,,
  \ee
 and the apparent luminosity:
 \bb
 \ell=\,\frac{L}{32 \pi }\,\frac{\hspace{14mm}W_e^5\hspace{4mm}(1+s^2)}{a_e^2c_e\,a_0^2c_0\,W_0\,(1+\sin x_0)}\,\Big| 1-{\textstyle\frac{1}{2}}\,\frac{I_{3\,0}}{f_0V_0}\Big|^{-1}.  
 \ee 
  Note that the initial condition $y_e$ drops out from the redshift and the apparent luminosity. Therefore only  the initial condition $s$ remains to be interpreted. It is related to the angle between the direction $-\vec x'(t_0)$  here and now pointing to the supernova at the north pole and the privileged direction of the axial Bianchi IX universe, the fourth Killing vector $\pa_z$. Let us denote this angle by $\gamma_0 $. He have: 
\bb
\cos \gamma = \,\frac{-\vec x'\cdot\vec\pa_z}{\sqrt{\vec\pa_z\cdot\vec\pa_z}}\, =\,\frac{s}{\sqrt{\,\frac{c^2}{a^2}\,+s^2} }\, \qq {\rm or} \qq
s=\,\frac{c}{a}\, \cot\gamma =\,\frac{c_0}{a_0}\, \cot\gamma_0 .
\ee

\subsection{Einstein's equations}

To write down the non-vanishing components of the Einstein tensor we will use the notations $H\dpp=a'/a$ and $H_c\dpp=c'/c$,
\bb
G_{tt}&=&3H^2+2H(H_c-H)+\,\frac{1}{a^2}\lp 4-\,\frac{c^2}{a^2}\rp,
\\[2mm]
G_{xx}&=&-\,\frac{a^2}{4}\lb H'+H'_c+H^2+H_c^2+HH_c+\,\frac{c^2}{a^4}\rb\, ,   
\\[2mm]
G_{zz}&=&-\,\frac{c^2}{4}\lb 2\,H'+3\,H^2+\,\frac{1}{a^2}\lp 4-3\,\frac{c^2}{a^2}\rp\rb ,   
\\[3mm]
&&G_{yy}=\cos^2x\,G_{xx}+\sin^2x\,G_{zz},\qq G_{yz}=\sin x\,G_{zz}.
\ee
Therefore we may assume that the energy-momentum tensor has only one non-vanishing component, the mass density $T_{tt}=\rho (t)$,  
`matter is comoving dust'. The covariant conservation of energy-momentum yields mass conservation:
\bb
\rho(t)=a_0^2c_0\,\frac{\rho _0}{a(t)^2c(t)}\, ,\qq
a_0\dpp = a(t_0),\ c_0\dpp = c(t_0),\ \rho _0\dpp =\rho(t_0).
\ee
The Einstein equations read:
\bb
3H^2+2H(H_c-H)+\,\frac{1}{a^2}\lp 4-\,\frac{c^2}{a^2}\rp&=&\Lambda +8\pi \,G\rho \label{tt},\\[2mm]
H'+H'_c+H^2+H_c^2+HH_c+\,\frac{c^2}{a^4}&=& \Lambda ,\label{xx9}\\[2mm]
 2\,H'+3\,H^2+\,\frac{1}{a^2}\lp 4-3\,\frac{c^2}{a^2}\rp  &=&  \Lambda .\label{zz}
\ee
In Bianchi I universes the $zz$ component (\ref{zz1}) of the  Einstein equation does not depend on the scale factor $c$.
 In Bianchi IX universes it does, (\ref{zz}). Therefore we use the
mass conservation to reduce the three Einstein equations to the $tt$ component and to the $xx$ component minus the $zz$ component.  We obtain the equivalent system:
 \bb
 3H^2+2H(H_c-H)+\,\frac{1}{a^2}\lp 4-\,\frac{c^2}{a^2}\rp
&=&\Lambda +8\pi\,G\,\rho _0\,\frac{a_0^2c_0}{a^2c},\label{tt9}
\\[2mm]
(H_c-H)'+(2H+H_c)(H_c-H)-
\,\frac{4}{a^2}\lp 1-\,\frac{c^2}{a^2}\rp
&=&0 .\label{xx-zz}
 \ee

\section{Deforming Friedman universes with comoving dust}

The axial Bianchi I, V and IX universes can be viewed as deformations of the flat, spherical and pseudo-spherical Friedman universes reducing the number of Killing vectors from six to four. To simplify the calculations and motivated by the phenomenological success of the $\Lambda $CDM model we will take these deformations to be infinitesimal. But before doing so it is useful to study those 
 infinitesimal deformations of the three Friedman universes that do not spoil any of the symmetries.

Consider the scale factor as function of cosmic time $a_F(t)$ and its Hubble parameter $H_F\dpp ={a'_F}/{a_F}$. The first Friedman equation is the $tt$ component of the Einstein equation and can be written
\bb
 3\,H_F^2 +3\,\frac{\kappa }{a_F^2}\, 
&=&\Lambda +8\pi\,G\,\rho _{F0}\,\frac{a_{F0}^3}{a_F^3}, \label{fried}
\ee
with four parameters: 
\begin{itemize}\item
 the `initial' condition today $a_F(t_0)\dpp=a_{F0}$,\\ (O course we mean `final', because we want to compute the past.)
\item
 the cosmological constant $\Lambda $,
 \item
 $8\pi\,G\,\rho _{F0}$ with $G$ being Newton's constant and $\rho _{F0}$
   the mass density of dust today,
   \item
 $\kappa =0,\,1,\,-1$ for the 3-spaces with zero, positive and negative curvature.
 \end{itemize}
A fifth parameter is important:  today's Hubble parameter also more appropriately called Lema{\^i}tre-Hubble constant, $H_F(t_0)=\dpp H_{F0}$. Its absolute value is determined by the first Friedman equation evaluated today. At the same time the choice of the other four parameters must obey the inequality,
\bb
\Lambda +8\pi\,G\,\rho _{F0}-3\,\frac{\kappa }{a_{F0}^2}\,\ge 0.
\ee
 Note also that in the flat case $\kappa =0$, the initial condition $a_{F0}$ is not essential.
 
 Being a first order ordinary differential equation, the Friedman equation has a unique local solution once we have chosen the four parameters in compliance with the inequality. 
 
  If the universe is not static,  the first Friedman equation  (\ref{fried}) implies the second Friedman equation (the $xx$ component of the Einstein equation),
\bb
2\,H'_F+3\,H_F^2+\,\frac{\kappa }{a_F^2}\, &=&\Lambda,
\ee
and the mass conservation, $\rho_F '+3\,H_F\,\rho_F =0 $, is satisfied if we set
\bb 
\rho_F\dpp=\rho _{F0}\,\frac{a_{F0}^3}{a_F^3}.
\ee
For a static universe (Einstein) we must have $\kappa =1$ and $a_{F0}^{-2}=\Lambda =4\pi\,G\,\rho _{F0}$. It has seven Killing vectors, but it is unstable.
From now on we will ignore the static universe and use the standard dimensionless parameters:
 \bb
 \Omega _{m0}\dpp=\,\frac{8\pi\,G\,\rho _{F0}}{3H_{F0}^2}\,,\qq 
  \Omega _{\Lambda 0}\dpp=\,\frac{\Lambda }{3H_{F0}^2}\,,\qq 
   \Omega _{\kappa 0}\dpp=\,\frac{-\kappa }{a_{F0}^2H_{F0}^2}\,=
   1-\Omega _{m0}-\Omega _{\Lambda 0}.
 \ee 
 Then a big bang or a bounce occurs and we take it as origin of cosmic time $a_F(0)=0$ or $H_F(0)=0$. 

\subsection{Deformations}

We want to deform the scale factor infinitesimally into
\bb
a=a_F(1+\eta),\qq \eta(t)\ll1.
\ee
We suppose that the function $\eta(t)$ is sufficiently small to neglect its quadratic and higher powers and indicate the linear approximation by $\sim\,$, e.g.
\bb
H\dpp=\,\frac{a'}{a}\, \sim H_F+\eta'.
 \ee
 We also suppose that the deformed scale factor satisfies the first Friedman equation with mass density today $\rho _0$,
 \bb
 3\,H^2 +3\,\frac{\kappa }{a^2}\, 
&=&\Lambda +8\pi\,G\,\rho _{0}\,\frac{a_{0}^3}{a^3},\label{fried1a}
\ee
Then the mass density today is also deformed infinitesimally into   
\bb
\rho _0=\rho _{F0}(1+R_0),\qq |R_0|\ll1.
\ee
The value of $R_0$ follows from the first Friedman equation in $a$ (\ref{fried1a}) evaluated today:
\bb
R_0&\sim&\,\frac{6\,H_{F0}\,\eta'_0-6\kappa \eta_0/a_{F0}^2}{8\pi G\,\rho _{F0}}\, .\label{rhoFfake}
\ee
Using the second Friedman equation in $a$ to first order we obtain:
\bb
\eta''+3\,H_F\eta'-\,\frac{\kappa }{a_F^2}\,\eta\,\sim\,0, \label{defFried}
\ee 
with initial conditions today $\eta'(t_0)=\eta'_0$ and $\eta(t_0)=\eta_0$.
%Equivalently we could have obtained equation (\ref{defFried}) by
%multiplying the first Friedman equation by $a_F^3$, deriving with respect to time and eliminating $H'_F$ with, 
%\bb
%a_F^3H'_F=-4\pi G\rho_{F0}a_{F0}^3+\kappa a_F. \label{h'}
%\ee
 In the flat case we may choose $a_0=a_{F0}$ implying $\eta_0=0$. 

These infinitesimal deformations respect the cosmological principle. We call them `fake', because we are interested in the minimal breaking of the principle.

 %It will however be difficult to disentangle the infinitesimal transformations that induce the minimal symmetry breaking from the fake ones for the case of axial Bianchi V universes. 

Before linearizing the infinitesimal deformations in axial Bianchi IX universes, we summarize the results of the axial Bianchi~I universe for the purpose of a strategic orientation.

\section{Axial Bianchi I}

The Einstein equations with cosmological constant and comoving dust in the generic Bianchi I universe (three Killing vectors) have explicite solutions \cite{stv}. In the axial Bianchi I universe, thanks to the fourth Killing vector, these solutions simplify considerably and can be written down without resorting to linearization. 

\subsection{Einstein equations}

We equate the first two scale factors, $a(t)=b(t)$. With the Hubble parameters  $H\dpp=a'/a$ and $H_c\dpp=c'/c$, the Einstein equations read:
\bb
3H^2+2H(H_c-H)&=&\Lambda +8\pi\,G\,\rho ,
\\[2mm]
 H'+H'_c+H^2+H_c^2+HH_c&=&\Lambda ,   \label{xx}   
\\[2mm]
 2\,H'+3\,H^2&=&\Lambda .
 \ee
 We can trade the $xx$ component (\ref{xx}) for the mass conservation with initial conditions today $t=t_0$,
 \bb
\rho(t)=a_0^2c_0\,\frac{\rho _0}{a(t)^2c(t)}\, ,\qq
a_0\dpp = a(t_0),\ c_0\dpp = c(t_0),\ \rho _0\dpp =\rho(t_0),\label{masscon}
\ee
 and remain with only two equations:
 \bb
3H^2+2H(H_c-H)&=&\Lambda +8\pi\,G\,\rho _0\,\frac{a_0^2c_0}{a^2c},\label{tt1}
\\[2mm]
 2\,H'+3\,H^2&=&\Lambda .\label{zz1}
 \ee
 Note that  the $zz$ component (\ref{zz1}) does not depend on the scale factor $c$ and with the short-hand $u\dpp =\sqrt{3\Lambda }\,t$ we obtain the solutions
 \bb
 a&=&a_0\,\frac{(\cosh u-1)^{1/3}}{(\cosh u_0-1)^{1/3}}\,, \label{a}
 \\[2mm]
 c&=&c_0\,\frac{(\cosh u-1)^{1/3}}{(\cosh u_0-1)^{1/3}}\,
 \frac{1+\xi \coth{\textstyle\frac{u}{2}} }{1+\xi\coth{\textstyle\frac{u_0}{2}} }\,,\qq \xi\in\rr,\label{c}
 \ee 
 where we have set the origin of the time axis at the big bang. The age of the Bianchi I universe today, with mass density today $\rho _0$,  
 is computed from
\bb
 g\,\dpp =\,\frac{8\pi\,G\,\rho _0}{\Lambda }\,&=&
 \,\frac{2}{(\cosh u_0-1)(1+\xi\coth{\textstyle\frac{u_0}{2}})}\,. \label{def}
 \ee
  In the Bianchi I universes, the two initial conditions $a_0$ and $c_0$ are not essential, we may and will set $c_0=a_0=a_{F0}$ by rescaling the coordinates.
 
 \subsection{A change of variables}
With $a=a_F$, equation (\ref{a}), the relation between the Friedmanian mass density  today $\rho _{F0}$ and the mass density of the axial deformation is given  by:
\bb
8\pi\,G\,\rho _{F0}&\dpp =&8\pi\,G\,\rho _{0}-2H_0\,(H_{c0}-H_0).\label{rhoF}
\ee
Let us replace the variable $c(t)$ in favor of the new {\it finite} variable $\eta(t)$:
 \bb
 c =\dpp a\,[1+\eta],\qq \eta>-1,\qq \eta_0=0.
 \ee
Then with
\bb 
3H^2=\Lambda +8\pi\,G\,\rho _{F0}\,\frac{a_0^3}{a^3},\qq\text{and}\qq
H_c-H=\,\frac{\eta'}{1+\eta}\,, 
\ee
 the $tt$ component of the Einstein equations (\ref{tt1}) becomes:
\bb
0\,=\,3H^2-\Lambda -8\pi\,G\,\rho _{F0}\,\frac{a_0^3}{a^3}&=&
8\pi\,G\,\rho _{F0}\,\frac{a_0^3}{a^3}\,\frac{-\eta}{1+\eta}\, +2H_0\,
\frac{a_0^3}{a^3}\,\frac{\eta'_0}{1+\eta}\,  
-2H\,\frac{\eta'}{1+\eta}\,,
\ee
or
\bb
 2\,\frac{a^3}{a_0^3}\,H{\eta'}+8\pi\,G\,\rho _{0}\,{\eta}-
2H_0\,{\eta'_0}(1+\eta)=0. \label{indeed}
\ee
With the solutions (\ref{a}) and (\ref{c}) we obtain:
\bb
\eta&=&\xi\,\frac{\coth{\textstyle\frac{u}{2}}-\coth{\textstyle\frac{u_0}{2}}}{1+\xi\coth{\textstyle\frac{u_0}{2}}}\, ,
\\[2mm]
\eta'&=&-\xi\,\frac{\sqrt{3\Lambda }}{(\cosh u-1)(1+\xi \coth{\textstyle\frac{u_0}{2}})}\, , \label{acc1}
\ee
which indeed solves equation (\ref{indeed}).
The parameter $\xi $ is related to $\eta'_0$ by
 \cite{sign}:
\bb
\xi &=&-2\,\frac{\eta'_0}{\sqrt{3\Lambda}\,g} \,.\label{acc2} 
\ee
For $\xi =0$, we have $\eta'_0=0$. Then $\eta$ vanishes identically and we are back at Friedman's solution.

Here we have used the change of variables $a=a_F$ and $c=a_F\,[1+\eta]$, which implied $\rho _0\not=\rho _{F0}$. This change of variables  is useful in Bianchi I, because the $zz$ component of the Einstein equation (\ref{zz1}) does not depend on the scale factor $c$. This is not true in the case of axial Bianchi IX and there, a more general change of variables will be mandatory, because the simpler one together with
 Einstein's equations implies isotropy, $\eta=0$.

 However these more general deformations will also contain fake ones.

\subsection{A more general change of variables and linearization}

The flat Friedman equations have the solution (\ref{a}).  Using the definition  (\ref{def}) with $\xi =0$ this solution takes the form:
 \bb
 a_F &=&a_{F0}\lp\frac{g_F}{2}\rp^{1/3}\lp\cosh u-1\rp ^{1/3},\qq g_F\,\dpp =\,\frac{8\pi\,G\,\rho _{F0}}{\Lambda }\,.
 \ee
 We define the new, more general, change of variables with two functions $\eta(t)$ and $\beta(t)$. For convenience we now take them infinitesimal and indicate the linear approximation by $\sim$.
\bb
a=\dpp a_F\,[1-\beta + \eta],&&H \sim H_F-\beta ' +\eta',
\label{new1}\\[2mm]
c=\dpp a_F\,[1+ \eta],\ \,&&H_c \sim H_F+ \eta'\,,
\qq\qq H_c-H\sim \beta ',
\label{new2}\\[2mm]
\rho _0=\dpp\rho _{F0}\,[1+R_0], &&|R_0|\ll 1.
\label{new3}
\ee
The initial conditions $c_0=a_0=a_{F0}$ and $a'_0$, $c'_0$ translate into
$\beta _0=\eta_0=0$ and $\beta'_0$, $\eta'_0$.

Note that if $\beta $ vanishes identically, the change of variables  is fake. For $\beta =\eta$ we recover the change of variables of the last subsection.

Now that we do not identify $a$ and $a_F$ anymore, it will be convenient  to keep in our trade of one Einstein equation for the mass conservation -- instead of the $zz$ component of the Einstein equation --  the  $xx$ component minus the $zz$ component:
\bb
(H_c-H)'+(2H+H_c)(H_c-H)=0,\label{x-z}
\ee
which integrates immediately:
\bb
H_c-H=(H_{c0}-H_0)\,\frac{a_0^2c_0}{a^2c}\,.
\ee 
Replacing $a$ and $c$ by $\beta $ and $\eta$ yields in linear order:
\bb
\beta ''+3H_F\,\beta '\sim 0,\qq \text{or}\qq
\beta ' \sim\beta '_0\,\frac{a_{F0}^3}{a_{F}^3}\, . \label{x-z1lin}
\ee
Replacing $a$ and $c$ by $\beta $ and $\eta$ in the $tt$ component (\ref{tt1}) and evaluating it today, we obtain:
\bb
R_0\sim\,\frac{4H_{F0}}{8\pi G\rho_{F0}}\,\alpha _0',\qq  \alpha \dpp=-\beta +{\textstyle\frac{3}{2}} \eta.
\ee
This result agrees with equation (\ref{rhoFfake}) for a fake deformation $\beta =0$, $\alpha ={\textstyle\frac{3}{2}} \eta$ in the flat case $\kappa =0$
and it
agrees with equation (\ref{rhoF}) for the special change of variables $\beta =\eta$, $\alpha= {\textstyle\frac{1}{2}} \eta$ in the last subsection.

Linearizing the $zz$ component of Einstein's equations (\ref{zz1}) we obtain:
\bb
\alpha ''+3H_F\, \alpha '\sim 0.
\ee
This result agrees with equation (\ref{defFried}) for a fake deformation $\beta =0$, $\alpha ={\textstyle\frac{3}{2}} \eta$ in the flat case $\kappa =0$.
Fake and non-fake variables decouple and we may set $\alpha =0 $ identically implying $\beta ={\textstyle\frac{3}{2}} \eta$, $a=a_F\,[1-{\textstyle\frac{1}{2}}  \eta],\,c= a_F\,[1+ \eta],\,\rho _0=\rho _{F0}$ and
\bb 
\eta'&\sim&\eta'_0\,\frac{a_{F0}^3}{a_F^3}\,, \label{eta1}
\ee
in accordance with the expression for $\eta$ in the old variables, equations (\ref{acc1}) and (\ref{acc2}). We are left with only one essential initial condition, $\eta'_0$, the `Hubble stretch'.

\section{Linearizing axial Bianchi~IX\label{b9l}}
\subsection{Linearizing Einstein's equations}

With our change of variables,
\bb
a=\dpp a_F\,[1-\beta + \eta],\qq c=\dpp a_F\,[1+ \eta],\qq
\rho _0=\dpp\rho _{F0}\,[1+R_0],
\label{new}
\ee
we linearize the last Einstein equation (\ref{xx-zz}):
\bb
\beta''+3\,H_F\,\beta'+8\,\frac{\beta}{a_F^2}\, &\sim& 0, \qq \text{with initial conditions}\ \beta_0\ \text{and}\ \beta'_0.
\ee
Linearizing the $tt$ component (\ref{tt9}) and evaluating it today, we obtain:
\bb
R_0\sim\,\frac{4H_{F0}\,\alpha _0'-4/a_{F0}^2\,\alpha _0}{8\pi G\rho_{F0}}\,,\qq \text{recall:\ }  \alpha \dpp=-\beta +{\textstyle\frac{3}{2}} \eta.
\ee
This result agrees with equation (\ref{rhoFfake}) for a fake deformation $\beta =0$, $\alpha ={\textstyle\frac{3}{2}} \eta$ in the spherical case $\kappa =1$.

Linearizing $2\cdot(\ref{xx9})+(\ref{zz})$
we obtain:
\bb
\alpha ''+3H_F\, \alpha '-\,\frac{\alpha }{a_F^2}\,\sim 0.
\ee
Note that this equation coincides with equation (\ref{defFried}) characterizing a fake deformation in the spherical case, $\kappa =1$.
Therefore -- like in the Bianchi I case -- fake and non-fake variables decouple, we may set $\alpha =0 $ identically and remain with $\beta ={\textstyle\frac{3}{2}} \eta$, 
\bb
a=a_F\,[1-{\textstyle\frac{1}{2}}  \eta],\qq
c= a_F\,[1+ \eta],\qq\qq\qq
\rho _0=\rho _{F0} \label{lin:a&c}
\ee
 and 
\bb
\eta''+3\,H_F\,\eta'+8\,\frac{\eta}{a_F^2}\, &\sim& 0. \label{eta9}
\ee
Note that the function $\eta$ is the Klein-Gordon field in \cite{gpk} and our equation (\ref{eta9}) agrees with equation (3.6) there.

Our four essential initial conditions, $a_0,\,c_0,H_0,\,H_{c0}$, fix $\rho _0$ by the $tt$ component of the Einstein equation (\ref{tt9}). They translate into the four essential initial conditions:
\bb 
a_{F0}={\textstyle\frac{2}{3}}a_0+{\textstyle\frac{1}{3}}c_0,
&&
H_{F0}\sim {\textstyle\frac{2}{3}} H_0+{\textstyle\frac{1}{3}} H_{c0},
\\[2mm]
\eta_0=\,\frac{c_0-a_0}{a_0+{\textstyle\frac{1}{2}} c_0}\,,
&&
\eta'_0\sim {\textstyle\frac{3}{2}} (H_{c0}-H_0).
\ee 
We call the last two `ellipticity' and `Hubble stretch'.
Being general properties of our change of variables (\ref{new1}-\ref{new3}) with $\alpha =0$, these four relations as well as $\rho _{F0}\sim\rho _0$ remain true at all times $t$. 

\subsection{Linearizing redshift and apparent luminosity} 

We start by linearizing $W(t)$ defined by the second of equations (\ref{4defs}),
\bb
W\sim\,\frac{2}{\sqrt{1+s^2}}\,a_F\lb1-{\textstyle\frac{1}{2}} \,\frac{1-2s^2}{1+s^2}\,\eta\rb,  \label{linW}
\ee
from which we obtain the linearized redshift,
\bb
\ul z+1\sim\,\frac{a_{F0}}{a_{Fe}}\,\lb 1- {\textstyle\frac{1}{2}} \,\frac{1-2s^2}{1+s^2}\,(\eta_0-\eta_e)\rb.
\ee 
Next we linearize the integral $I_3(t)$, equation (\ref{I3}),
\bb
I_3\sim \,\frac{6}{(1+s^2)^{3/2}}\,\int_{t_e}^t\,\frac{\eta}{a_F}\,,
 \ee
 and linearize  the apparent luminosity:
 \bb
   \ell\sim\,\frac{L}{2 \pi }\,\frac{a_{Fe}^2}{a_{F0}^4}\,\frac{1}{(1+s^2)(1+\sin x_0)} \lb 1+
 {\textstyle\frac{1}{2}} \frac{1 -2s^2}{1+s^2}\,(\eta_0-5\eta_e)
 +
  \frac{3}{(1+s^2)^{3/2}}\,\frac{\chi _{e0}\bar\eta_{e0}}{f_0V_0}    \rb,  
 \ee 
 with the definitions:
\bb
\chi_e (t)\dpp =\int_{t_e}^t \,\frac{1}{a_F}\,,\qq
\chi_{e0}\dpp =\chi_e(t_0),
\label{defchie}\\ 
\bar\eta_e(t)\dpp =\,\frac{1}{\chi_e (t)}\,  \int_{t_e}^{t} \,\frac{\eta}{a_F}\,,\qq 
\bar\eta_{e0}\dpp=\bar\eta_e(t_0).
\ee
Remember that we used $\chi $ for the radial coordinate  in equation (\ref{chi}). This $\chi $ is the geodesic distance from the north pole on the 3-sphere. We have set the speed of light to unit. Therefore the conformal time $\chi _e(t)$ coincides with the comoving geodesic distance of our photon emitted at the north pole $\chi =0$ at cosmic time $t_e$ in the underlying Friedman universe, $\eta=0$.

To compute the apparent luminosity of our photon in the axial Bianchi IX universe in linear approximation, we still need $f(x)\,V(x)$ to zeroth order in $x-x_F$ and $1+\sin x$ to first order. To this end we linearize equation (\ref{firstin}) and obtain:
\bb
f(x_F)\,V(x_F)&=&\frac{\tan\chi _e}{\sqrt{1+s^2}}\,,\\
1+\sin x&\sim& 
\frac{2}{1+s^2}\, \sin^2\chi_e  
 \lb
1+\frac{1+4s^2}{1+s^2} \,\frac{\chi _e\,\bar\eta_e }{\tan \chi_e }\rb.
\ee 
Now, the apparent luminosity takes the form:
\bb
     \ell\sim \ell_F
\lb 1+ {\textstyle\frac{1}{2}} \,\frac{1 -2s^2}{1+s^2}\lp
 \eta_0-5\eta_e+4\,\frac{\chi _{e0}}{\tan\chi_{e0} }\,\bar\eta_{e0} \rp   \rb 
\ee
with
\bb
\ell_F=\,\frac{L}{4\pi\,a_{F0}^2\,\sin^2\chi_{e0} }\lp\frac{a_{Fe}}{a_{F0}}\rp^2.
\ee
Note that the initial condition $s$ appears in the linearized redshift $z$ (dropping the underline) and in the linearized apparent luminosity always multiplied by $\eta$. Therefore to first order, we may replace $s$ here simply by $s\sim\cot \gamma _0$. Note that in this approximation the angle $\gamma $ also coincides with the coordinate $\theta $, which is constant for geodesics coming from the north pole. Therefore:
\bb
z+1
&\sim&\,\frac{a_{F0}}{a_{Fe}}\,\lb 1- \,\frac{1-3\cos^2\theta}{2}\, (\eta_0-\eta_e)\rb, 
\label{redshift2}\\[2mm]
\ell&\sim& \ell_F
\lb 1+\,\frac{1-3\cos^2\theta}{2}\,\lp
 \eta_0-5\eta_e+4\,\frac{\chi _{e0}}{\tan\chi_{e0} }\,\bar\eta_{e0} \rp    \rb. 
 \label{apLum}
\ee
In the limit of zero curvature, $a_0$ and $c_0\rightarrow\infty$,
 we have 
\bb
\lim\sin\chi_{e0} =\chi_{e0},\qq
\lim\,\frac{\chi _{e0}}{\tan\chi _{e0}}\,  =1,
\ee
 and the differential equation for $\eta$ in Bianchi IX (\ref{eta9}) reduces to the one in Bianchi~I (\ref{eta1}). Taking due account of our change of definitions, $\eta_{\rm I}= {\textstyle\frac{3}{2}}\eta_{\rm IX}  $ and $ \chi_{\rm I}=\int 1/a$ while here $ \chi_{\rm IX}=\int 1/a_F$, we reproduce our perturbative results \cite{stv} of axial Bianchi I.
 
 Note the $\cos^2\theta$  in equations (\ref{redshift2}) and (\ref{apLum}) ensuring  that the Lema{\^i}tre-Hubble diagram in Bianchi I and IX universes has forward-backward symmetric anisotropies (quadrupole). They can be  distinguished easily from the anisotropies induced by a peculiar velocity of the observer (the kinetic dipole), that do not share this symmetry.

\section{Exact solutions of Friedman's equations 
\\ ${}$ \hspace{79mm}
with comoving dust
 \label{exact}}

To speed up the fitting of axial Bianchi IX universes 
with the  Lema{\^i}tre-Hubble diagram,
 it will be useful to have exact solutions of the Friedman equations (\ref{fried}).

Exact solutions of the Friedman equations in presence of a cosmological constant, curvature, dust and radiation are well-known, for example in terms of Weierstra{\ss}  elliptic functions \cite{lemaitre}, \cite{exact4}-\cite{exact6}.

We will not have to include radiation, which considerably simplifies the solutions and we will use Jacobi elliptic functions, following Edwards \cite{edw}.

\subsection{Jacobi elliptic and related functions \label{subsub}}

 The Jacobi elliptic functions $\cn(u,k^2),\,\sn(u,k^2),\,\dn(u,k^2)$ with $k^2\in[0,1]$ interpolate between trigonometric functions for $k^2=0$ and hyperbolic ones for $k^2=1$ according to 
\bb
\cn(u,0)=\cos u,\ & \sn(u,0)=\sin u,\qq\ \ &\dn(u,0)=1,\\[2mm]
\cn(u,1)=\tanh u,&\sn(u,1)={1}/{\cosh u}, &\dn(u,1)={1}/{\cosh u}.
\ee
These functions may be defined from the differential system (dropping the second argument, which will always be $k^2$):
\bb
\,\frac{\de}{\de u}\,\cn u=- \sn u\,\dn u,\qq
\,\frac{\de}{\de u}\,\sn u= \cn u\,\dn u,\qq
\,\frac{\de}{ \de u}\,\dn u=-k^2 \cn u\,\sn u,\label{3deriv}
\ee
with the initial conditions
\[\sn(0)=0,\qquad\qquad \cn(0)=1, \qquad\qquad \dn(0)=1.\]
These definitions imply the algebraic relations
\bb\label{alg}
\cn^2 u+\sn^2 u=1,\qq \dn^2 u+k^2\sn^2 u=1.
\ee
We will indicate the inverse functions by arc. 
Of paramount importance are the elliptic integrals $K$ and $K'$ (not a derivative) defined by 
\bb\label{KKp}
K(k^2)\dpp=\int_0^{\pi/2}\frac{\de x}{\sqrt{(1-x^2)(1-k^2\,x^2)}}, \qquad\qquad K'(k^2)\dpp=K(1-k^2). 
\ee
In fact Jacobi even defined the functions $\cn(u)$, $\sn(u)$ and $\dn(u)$   for complex values of $u$, revealing that their squares are meromorphic and doubly periodic functions with periods $(2K,\,2iK')$. What is special with elliptic functions is that besides their double periodicity they must have either a double pole or two simple poles in their period parallelogram... otherwise they are reduced to a constant.

\subsection{The scale factor as function of conformal time}

Let us change the independent variable from cosmic time $t$ to conformal time  = comoving geodesic distance, 
\bb
\chi (t)\dpp=\int_0^t\frac{1}{a_F},\qq 0\leq t.
\ee 
We will set the big bang at $t=0$, $a_F(0)=0$, but the corresponding singularity will be integrable. 
Note the link with our former definition (\ref{defchie}):
\bb
\chi _e(t)=\chi (t)-\chi (t_e).
\ee

Let us also change the dependent variable from $a_F$ to $s$ (which of course has nothing to do with the initial condition $s$ in sections \ref{b9} and \ref{b9l}),
\bb
s(\chi)\dpp= \,\frac{1}{a_{F0}}\,a_F(t(\chi )), 
\ee
where by abuse of notations we write the inverse function of $\chi (t)$ as 
$t(\chi )$. 

Let us also change dimensionless parameters assuming that the cosmological constant is positive:
\bb
\Lambda  >0,\qq\qq
\gamma \dpp=4\,\frac{\Omega _{m 0}}{\Omega _{\Lambda 0}}\, ,\qq
\nu\dpp=-\,\frac{4}{3}\, \frac{\Omega _{\kappa  0}}{\Omega _{\Lambda 0}}\, .
\ee
Note that $\nu$ and $\kappa $ carry the same signs or vanish both.

Then Friedman's equation (\ref{fried}), retaining only the positive root, becomes:
\bb
a_{F0}\sqrt{\frac{\Lambda}{3}}d\chi=\frac{ds}{\sqrt{s\,
(s^3-{\textstyle\frac{3}{4}}\nu\,  s+{\textstyle\frac{1}{4}} \gamma )}}. \label{fried2}
\ee 
Let us assume that
\bb
\gamma ^2-\nu^3\,=\,16\,\frac{\Omega _{m 0}^2}{\Omega _{\Lambda 0}^2}
\lp1+\,\frac{4}{27}\, \frac{\Omega _{\kappa  0}^3}{\Omega _{\Lambda 0}\Omega _{m 0}^2}\rp >0,
\ee
is positive. This implies that there is a big bang. If $\gamma ^2-\nu^3$ vanishes we have Einstein's static universe with seven Killing vectors. Universes with negative $\gamma ^2-\nu^3$ are spherical and undergo one bounce. They contain the maximally symmetric de Sitter spacetimes (positive curvature, ten Killing vectors). Nevertheless  we will ignore this last class, because it is frankly incompatible with today's observations, which we would like to try and fit in section \ref{Andre} with universes  breaking minimally the isometry group in the spaces of simultaneity  (positive curvature, four Killing vectors). 

Since $\gamma ^2-\nu^3$ is positive, the cubic polynomial  on the right-hand side of equation (\ref{fried2}) has one real root given by Cardano's formula 
\bb 
\tilde s= -{\textstyle\frac{1}{2}} \lp\gamma +\sqrt{\gamma ^2-\nu^3}\rp^{1/3}
-{\textstyle\frac{1}{2}} \lp\gamma -\sqrt{\gamma ^2-\nu^3}\rp^{1/3},
\ee
and two complex conjugate roots $s_1\pm i\,s_2$, leading to the factorization
\bb\label{fact}
s^3-{\textstyle\frac{3}{4}}\nu\,  s+{\textstyle\frac{1}{4}} \gamma =(s-\tilde s)\Big((s-s_1)^2+s_2^2\Big),
\qq\qq s_1=-\frac{\tilde s}{2},\qq\qq\qq s_2^2=\frac 34(\tilde s^2-\nu)>0.
\ee
The last inequality allows us to define three dimensionless positive constants
\bb \label{pos}
A\dpp=\sqrt{\tilde s^2-{\textstyle\frac{3}{4}} \nu}\qq<\qq
B\dpp=\sqrt{3\tilde s^2-{\textstyle\frac{3}{4}} \nu},\qq\qq 
k^2\dpp=\,\frac{1}{2}\, +\,\frac{3}{4}\,\frac{\tilde s^2-\nu/2}{AB},
\ee
and to check that $k^2<1$ since we have
\bb
k^2(1-k^2)=\frac{4A^2B^2-9(\tilde s^2-\nu/2)^2}{16A^2B^2}=3\,\frac{\tilde s^2(\tilde s^2-\nu)}{16A^2B^2}>0.
\ee
(Of course the constants $A$ and $B$ have nothing to do with the conserved quantities defined in equations (\ref{conserved1}) and (\ref{conserved3}).)

We are now in position to express the scale factor in terms of the conformal time: following reference \cite{bf} we operate a second change of dependent variable $s(\chi )\rightarrow u(\chi )$ defined by
\bb
s=|\tilde s|A\,\frac{1-\cn u,}{(A+B)\cn u-(A-B)}\qq 
\Longleftrightarrow \qq \cn u=\,\frac{(A-B)s-A\tilde s}{(A+B)s-A\tilde s}.\label{doublearrow}
\ee
The  {\it raison d'\^etre} of this change of variables is the following equation:
\bb\label{trans}
\frac{\de s}{\sqrt{s\,
(s-\tilde s)((s-s_1)^2+s_2^2)}}\,  
=\,\frac{\de u}{\sqrt{AB}},\qquad s\geq 0.\label{toprove}
\ee
{\bf Proof:}
Using the relations (\ref{alg}), and the specific form of $k^2$ given in (\ref{pos}), a purely algebraic computation gives
\bb
\sn u=2\sqrt{AB}\,\frac{\sqrt{s(s-\tilde s)}}{(A+B)s-A\tilde s},\qquad \qquad \dn u=|\tilde s|\frac{\sqrt{(s-s_1)^2+s_2^2}}{(A+B)s-A\tilde s}.
\ee
Differentiating the 
second of equations (\ref{doublearrow}) with respect to $s$
 gives
\bb
\de s=\frac{((A+B)s-A\tilde s)^2}{2AB|\tilde s|}\,\sn u\,\dn u\,\de u.
\ee
These last three relations do imply equation (\ref{trans}). $\blacksquare$

So we can write the Friedman equation (\ref{fried2}) conveniently as  
\bb
\de u=\sigma\,\de\chi, \qq\qq \sigma \dpp=a_{F0}\sqrt{\frac{\Lambda\,AB}{3}},
\ee
which leads to $u=\sigma\,\chi $ (setting the big bang at  $\chi=0$) and the first of equations (\ref{doublearrow}) yields  the scale factor as function of conformal time:
\bb
\boxed{
a_F(t(\chi ))=a_{F0}|\tilde s|\,\frac{A}{A+B}\,\frac{1-\cn(\sigma \chi )}{\cn(\sigma \chi )-\cn(\sigma \chic )}\,,
}
\qquad \cn (\sigma\chic)\dpp=\frac{A-B}{A+B}\,.  \label{rr1}
\ee
The big bang takes place at $\chi =0$, $t(0)=0$, because cn$\,0=1$.
The age of the universe in conformal time is:
\bb\label{conf}
\chi _0\,=\,\frac{1}{\sigma }\,\,\,  \text{arc}\,\cn\,\frac{A\,(1+|\tilde s|)-B}{A\,(1+|\tilde s|)+B}\,.
\ee
Note that at finite conformal time $\chic$ the scale factor tends to infinity. The corresponding cosmic time $t(\chic)$ is infinite, indeed the integral
\bb
 \int_0^\infty\,\frac{\de t}{a_F(t)}\, =\,\chic
\ee
is convergent. 

The relation (\ref{rr1}) was first given by Edwards \cite{edw}, but for the reader's ease we have given the computational details.

\subsection{Cosmic time as function of conformal time}
Here we need two Theta functions.  
We follow the conventions of reference \cite{ww},
\bb\barr{l}\dst 
\theta _1(v,q)\dpp=2\sum_{n\geq 0}(-1)^n\,q^{(n+1/2)^2}\,\sin((2n+1)v),\\[5mm] \dst 
\tht_4(v,q)\dpp=1+2\sum_{n\geq 1}(-1)^n\,q^{n^2}\,\cos(2nv),\earr 
\ee
and in Jacobi's notations
\bb
q\dpp=e^{-\pi K'/K},\qq H(u)\dpp=\tht_1\left(\frac{\pi u}{2K},q\right),
\qq  \Theta(u)\dpp=\tht_4\left(\frac{\pi u}{2K},q\right),\qq \zeta(u)\dpp=\,\frac{\de}{\de u}\, \ln H(u) . 
\ee
(The function $H(u)$ has nothing to do with the Hubble parameter.)

Then cosmic time as a function of conformal time is
\bb
\boxed{
t(\chi )=\sqrt{\frac{3}{\Lambda }}\la\!{\textstyle\frac{1}{2}} \ln\!\!\lp\!
\frac{H(\sigma (\chi +\chic))}{H(\sigma (-\chi +\chic))}
\,\frac{\widehat{s}\,\,\dn(\sigma \chi )+\widehat{d}\,\, \sn(\sigma \chi )}{\widehat{s}\,\,\dn(\sigma \chi )-\widehat{d}\,\,\sn(\sigma \chi )\!}\rp
\!\!-\!\!\lp \!\!\zeta(\sigma \chic)   +\frac{\widehat{d}}{\widehat{s}}\,\!\rp\!\sigma \chi \ra ,
}
\label{rr2}
\ee
where we use the abbreviations:
\bb\label{hat}
\wh{s}\dpp=\sn\wh{u}=2\,\frac{\sqrt{AB}}{A+B},\qq\qq \wh{c}\dpp=\cn\wh{u}= \frac{A-B}{A+B},\qq\qq \wh{d}\dpp=\dn\wh{u}=\frac{|\tilde s|}{A+B},\qq \wh{u}=\si\wh{\chi}.
\ee
{\bf Proof:} Our starting point is
\bb
\de t=a_F\,\de\chi=\frac{a_F}{\sigma}\,\de u=\frac{a_{F0}|\tilde s|A}{\si(A+B)}\,\frac{1-\cn u}{\cn u-\wh{c}}\,\de u. 
\ee
Let us write
\bb
\frac{1-\cn u}{\cn u-\wh{c}}=\frac{1-\wh{c}}{\wh{s}\,\wh{d}}\left(-\frac{\wh{s}\,\wh{c}\,\wh{d}}{\sn ^2 u-\wh{s}^2}-\frac{\wh{s}\,\wh{d}}{1-\wh{c}}-\wh{s}\,\wh{d}\,\frac{\cn u}{\sn ^2 u-\wh{s}^2}\right).
\ee
The relations (\ref{hat}) impy that
\bb
\frac{a_{F0}|\tilde s|A}{\si(A+B)}\frac{1-\wh{c}}{\wh{s}\,\wh{d}}=\sqrt{\frac 3{\Lambda}},
\ee
and we are led to
\bb\label{dtdu}
\frac{\de t}{\de u}=\sqrt{\frac 3{\Lambda}}\left(-\frac{\wh{s}\,\wh{c}\,\wh{d}}{\sn ^2 u-\wh{s}^2}-\frac{\wh{s}\,\wh{d}}{1-\wh{c}}-\wh{s}\,\wh{d}\,\frac{\cn u}{\sn ^2 u-\wh{s}^2}\right).
\ee
Let us check this relation 
 by deriving the right-hand side of equation (\ref{rr2}) with respect to $u$.
 First, using the derivatives  (\ref{3deriv}) and the algebraic relations (\ref{alg}), one gets
\bb
\frac \de{\de u}\left(\frac 12\,\ln\,\frac{\widehat{s}\,\,\dn u+\widehat{d}\,\, \sn u}{\widehat{s}\,\,\dn u-\widehat{d}\,\,\sn u\!}\right)=-\wh{s}\,\wh{d}\,\frac{\cn u}{\sn ^2 u-\wh{s}^2}.
\ee
For the remaining terms we need a relation between Theta functions given in \cite{ww}[p.487]
\bb
\tht_4^2(0)\tht_1(y+z)\,\tht_1(y-z)=\tht_1^2(y)\,\tht_4^2(z)-\tht_4^2(y)\,\tht_1^2(z),
\ee
or in Jacobi notations
\bb
\Theta^2(0)H(u+v)H(u-v)=H^2(u)\,\Theta^2(v)-\Theta^2(u)\,H^2(v).
\ee
Dividing by $H^2(u)\,H^2(v)$ and taking into account that
\bb
\sn u=\frac 1{\sqrt{k}}\,\frac{H(u)}{\Theta(u)}\,,
\ee
we obtain
\bb 
\Theta^2(0)\frac{H(u+v)H(u-v)}{H^2(u)\,H^2(v)}=\frac 1k\,\frac{\sn ^2 u-\sn ^2 v}{\sn^2 u\,\sn^2 v}.
\ee
Taking the logarithm and differentiating with respect to $u$ gives  
\bb
\frac 12\Big(\zeta(u+v)+\zeta(u-v)-2\zeta(u)\Big)=\frac{\cn u\,\dn u}{\sn u}\frac{\sn^2 v}{\sn^2 u-\sn^2 v}.
\ee
The substitutions $u\,\to\,\wh{u},\ v\,\to\, u$ lead to
\bb\label{zeta}
\frac 12\Big(\zeta(u+\wh{u})+\zeta(-u+\wh{u})-2\zeta(\wh{u})\Big)=-\frac{\wh{c}\,\wh{d}}{\wh{s}}\frac{\sn^2 u}{\sn^2 u-\wh{s}^2}=-\frac{\wh{s}\,\wh{c}\,\wh{d}}{\sn ^2 u-\wh{s}^2}-\frac{\wh{c}\,\wh{d}}{\wh{s}}.
\ee
Coming back to (\ref{rr2}) the leftover piece 
\bb
\frac 12\ln\frac{H(u+\wh{u})}{H(-u+\wh{u})}-\left(\zeta(\wh{u})+\frac{\wh{d}}{\wh{s}}\right)u,
\ee
when differentiated, gives
\bb 
\frac 12\Big(\zeta(u+\wh{u})+\zeta(-u+\wh{u})-2\zeta(\wh{u})\Big)-\frac{\wh{d}}{\wh{s}}
\ee
and thanks to (\ref{zeta}) we obtain for this piece
\bb
-\frac{\wh{s}\,\wh{c}\,\wh{d}}{\sn ^2 u-\wh{s}^2}-\frac{\wh{c}\,\wh{d}}{\wh{s}}-\frac{\wh{d}}{\wh{s}}=-\frac{\wh{s}\,\wh{c}\,\wh{d}}{\sn ^2 u-\wh{s}^2}-\frac{\wh{s}\,\wh{d}}{1-\wh{c}}.
\ee
This concludes the proof of (\ref{rr2}), since $t(0)=0.\qq\blacksquare$

\subsection{Checks}

We have checked numerically that in the flat case, $\nu=0$, the exact solution (\ref{rr1}) coincides with 
\bb
a_F(t(\chi ))=a_{F0}\,\frac{(\cosh [\sqrt{3\Lambda }\,t(\chi )]-1)^{1/3}}{(\cosh [\sqrt{3\Lambda }\,t(\chi _0)]-1)^{1/3}}\,
\ee
from equations (\ref{a}) and (\ref{rr2}).

    In the case with non-vanishing curvature, we checked the exact solution (\ref{rr1}) by integrating the Friedman equation with a Runge-Kutta method.

For the user's convenience table \ref{access} summarizes a few commands accessing Jacobi elliptic functions and others in Mathematica and Maple.

\begin{table}[!h]
\begin{center}
\begin{tabular}{|c||c|c|}
\hline
& Mathematica&Maple\\\hline\hline
$\cn(u,k^2)$&JacobiCN[u,k$\,\hat{}\,$2]&JacobiCN(u,k)\\\hline
arc$\,\cn(u,k^2)$&InverseJacobiCN[u,k$\,\hat{}\,$2]&InverseJacobiCN(u,k)\  \\\hline
$K(k^2)$&EllipticK[k$\,\hat{}\,$2]&EllipticK(k)\\\hline
$\theta _i(v,q),\ i=1,2,3,4$&EllipticTheta[$i$, v, q]&JacobiTheta$i$(v,q)\\\hline
\end{tabular}
\caption[]{A few access commands in Mathematica and Maple. Note the absence of the square on $k$ in Maple. Note also that EllipticTheta[$i$, v, q]  are not  elliptic  functions.}
 \label{access}
 \end{center}
\end{table}

\section{
The Lema{\^i}tre-Hubble diagram in  axial Bianchi IX \\
${}$ \hspace{49mm}
 universes with small anisotropies
 \label{Andre}}

The Lema{\^i}tre-Hubble diagram is the parametric plot $(\ell(t_e), z(t_e))$ whose parameter is the cosmic emission time $t_e$. As the light arriving in our telescope cannot tell us its emission time it is convenient to use as parameter the conformal emission time $\chi (t_e)$ instead of the cosmic emission time $t_e$. Then the fit with the supernova data follows the following steps:
\begin{enumerate}
\item
 Choose a spherical or flat Friedman universe with comoving dust characterized by three positive parameters: $H_{F0},\, \Omega _{\Lambda 0},\,\Omega _{m 0}$, such that 
 \bb
   \Omega _{\Lambda 0}+\Omega _{m0}\ge1 && \text{(positive or zero curvature, $\Omega _{\kappa  0}\le0$),}\\
   1+\,\frac{4}{27}\, \frac{\Omega _{\kappa  0}^3}{\Omega _{\Lambda 0}\Omega _{m 0}^2}>0 &&\text{(initial big bang)}.
   \ee
   Then the age of the universe in conformal time $\chi _0$ is given by equation (\ref{conf}) and in cosmic time $t_0=t(\chi _0)$ by equation (\ref{rr2}). 
The scale factor today  is 
   \bb 
   a_{F0}=\,\frac{1}{H_{F0} \sqrt{-\Omega _{\kappa  0}}}\, .
   \ee
   For  any conformal time $\chi $ the scale factor $a_\chi (\chi )\dpp=a_F(t(\chi ))$ is given by equation (\ref{rr1}). 
\item
Fix a direction $\pa_z$ privileged by the axial symmetry in the sky by two parameters, e.g. (right ascension,\,declination). 
\item
Choose two more {\it small} parameters: the final conditions of the infinitesimal anisotropy $\eta(t)$ today. Since we will work with the conformal time, let us define $\eta_\chi (\chi )\dpp=\eta(t(\chi ))$ and the final conditions today by:
\bb
\eta_\chi (\chi _0)=\eta(t_0)=\eta_0&&\text{`ellipticity'},\\[2mm]
\,\frac{\de}{\de \chi } \eta_\chi (\chi _0)=a_{F0}\,\frac{\de}{\de t}\eta(t_0)  &&\text{`Hubble stretch'}. \ee
Both should  be smaller in absolute value than 10 \%.
Note that in the flat case the ellipticity vanishes by the choice $a_0=c_0$ and Bianchi I only has the Hubble stretch.
\item
Rewriting equation (\ref{eta9}) in terms of conformal time, we have:
\bb
\,\frac{\de^2}{\de \chi ^2} \eta_\chi+
2\, H_\chi \,\frac{\de}{\de \chi } \eta_\chi
+
8\, \kappa \,\eta_\chi
\sim 0,\qq \kappa =1,\, 0,
\ee
with the conformal Hubble parameter
\bb
H_\chi \dpp=\frac{\de}{\de \chi } a_\chi\,/a_\chi = \si(1-\wh{c})\,\frac{\sn u\,\dn u}{(1-\cn u)(\cn u-\wh{c})}\,.\ee
Solve this linear differential equation numerically for times prior to $\chi _0$ with the final conditions: ellipticity and Hubble stretch.
\item
Choose a supernova with its redshift $z$, luminosity $\ell_{\it obs}$ and its direction. Let $\theta $ be the angle between this direction and the chosen direction privileged by the axial symmetry.

Invert numerically
\bb
z+1
&\sim&\,\frac{a_{F0}}{a_\chi(\chi _e)}\,\lb 1- \,\frac{1-3\cos^2\theta}{2}\, (\eta_0-\eta_\chi (\chi _e))\rb
\ee
to obtain the emission time $\chi _e\dpp =\chi(t_e)$ for a given redshift and angle $\theta $. Attention, the conformal emission time is a number and should not be confused with the function 
$\chi_e (t)\dpp =\int_{t_e}^t a_F^{-1}$, equation (\ref{defchie}). We have
$\chi _e(t_0)=\dpp\chi _{e0}=\chi _0-\chi _e$ and $\eta_\chi (\chi _e)=\eta(t_e)=\eta_e.$

For consistency of the linearization, $\eta_e$ must remain smaller in absolute value than\\ 10 \%.

\item
Evaluate numerically the integral
\bb
\bar\eta_{e0}\dpp=\,\frac{1}{\chi _{e0}}\int_{\chi _e}^{\chi _0}\eta_\chi .
\ee 
\item
Compute the theoretical apparent luminosity,
\bb
\ell\sim \,\frac{L}{4\pi\,a_{F0}^2\,\sin^2\chi_{e0}\,(z+1)^2 }\lb 1+\,\frac{1-3\cos^2\theta}{2}\,\lp
 \eta_0-5\eta_e+4\,\frac{\chi _{e0}}{\tan\chi_{e0} }\,\bar\eta_{e0} \rp    \rb, 
\ee
and compare it to the observed apparent luminosity $\ell_{\it obs}.$
\item
Let $\Delta_i $ be $\ell_{\it obs}-\ell$ for the $i$th supernova and let $V$ be the covariance matrix encoding the experimental error bars.  Define
$\chi ^2 \dpp=\sum_{ij} \Delta _iV^{-1}_{ij}\Delta _j.$ (Of course this $\chi ^2$ has nothing to do with conformal time.)
\item
Minimize $\chi ^2 $ over the set of the seven parameters:
\\ ${}$\hspace{4mm} 
$H_{F0},\, \Omega _{\Lambda 0},\,\Omega _{m 0}$, (right ascension,\,declination), ellipticity and Hubble stretch\\ in their domains of definition.
 \end{enumerate}
 
 Axial Bianchi I is the five-parameter sub-model with vanishing curvature, $\Omega _{\Lambda 0}+\Omega _{m 0}=1$, and vanishing ellipticity, $\eta_0=0$.
 
   Note that the direction dependence is captured by the quadrupole coefficient $(1-3\cos^2\theta)/2$ in both redshift and apparent luminosity as function of emission time. This coefficient varies from -1 for $\theta =0$ to 1/2 for $\theta =90^\circ$ and vanishes around $\theta =54.7^\circ$. In this direction the Lema{\^i}tre-Hubble diagrams of the axial Bianchi IX and Bianchi I universes coincide with the direction-independent
 Lema{\^i}tre-Hubble diagrams of their underlying spherical and flat Friedman universes.
 
 As an example we display in figure 2 the Lema{\^i}tre-Hubble diagrams $\ell(z)$ with $\theta =0$ simultaneously for the Bianchi IX universe with $\Omega _{\Lambda 0}=0.35$, $\Omega _{m 0}=0.75$, $\eta_0=0$, $\de/\de \chi \,\eta_\chi (\chi _0)=0.04$, for the Bianchi I universe with $\Omega _{\Lambda 0}=0.3$, $\Omega _{m 0}=0.7$, $\eta_0=0$, $\de/\de \chi \,\eta_\chi (\chi _0)=-0.01$ and their underlying Friedman universes. 
 
 The example suggests that small positive curvature and small anisotropies do co-exist peacefully in axial Bianchi IX universes during our recent past.
 
 \renewcommand{\thefigure}{2}
 \begin{figure}[htbp]
\includegraphics[width=8cm, height=4.8cm]{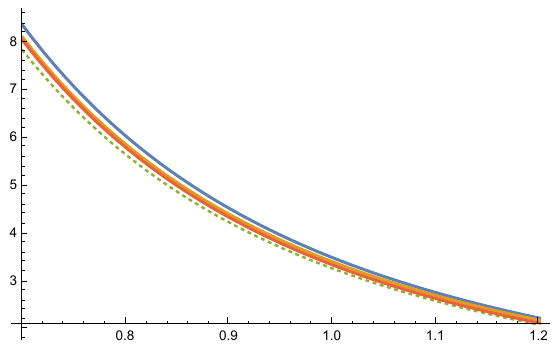}
\includegraphics[width=8cm, height=4.8cm]{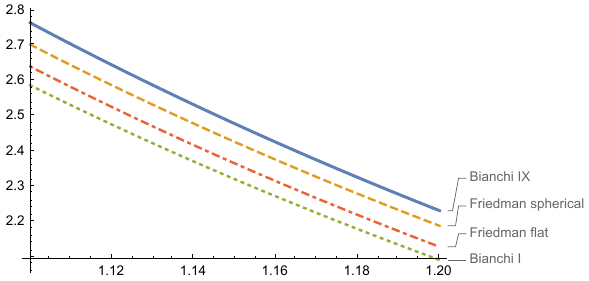}
\caption[]{apparent luminosity $\ell$ in arbitrary units versus redshift $z$ \\ 
${}$\hspace{19mm}
blue, continuous: Bianchi IX, \hspace{12mm}
       orange, dashed: Friedman spherical,\\
      ${}$ \hspace{23.5mm} 
       green, dotted: Bianchi I,
    \hspace{8mm}   red, dashed dotted: Friedman flat
}
\end{figure}
 
 In 2014 we confronted \cite{stv} this sub-model with the 740 type 1a supernovae below a redshift of $z=1.3$ in the Joint Light curve Analysis \cite{jla} and found only a 1-$\sigma $ signal. 
 
 Therefore we think that a fit of the seven-parameter axial Bianchi IX model to the Lema{\^i}tre-Hubble diagram of supernovae becomes reasonable only once we can include the data expected from the Vera Rubin Observatory.
 It should start operating in 2024 and observe 50\,000   supernovae per year.
 
 Complementarily, the James Webb Space Telescope is expected to observe some 200 type 1a supernovae up to redshift $z=6$ in the next six years \cite{jwst}. There is also the Chinese Space Station Telescope. It should start operating in 2024 and observe some 1800 type 1a supernovae below a redshift of $z=1.3$ in a time span of two years \cite{csst}.
 
Also in 2014 Cea \cite{cea1} fitted the axial Bianchi I model to the WMAP and Planck data at redshift 1090. He also found a 1-$\sigma $ signal. Although his Hubble stretch has opposite sign and is smaller than ours by eight orders of magnitude, our results are compatible with his.
 
 Again in 2014 Darling \cite{dar} used the tri-axial Bianchi I model (seven parameters) to fit the apparent motion of 429 extra-galactic radio sources measured by Titov \& Lambert \cite{tl}  
 using Very Long Baseline Interferometry.  His main Hubble stretch has the same sign as ours but is ten times larger and the results are again compatible statistically.
 
 Waiting and preparing for the mentioned supernovae to be observed, a combined analysis of axial Bianchi IX universes with the Lema{\^i}tre-Hubble diagram, Cosmic Microwave Background, drift of radio sources and Baryonic Acoustic Oscillations (and maybe with weak lensing or black-hole mergers) is called for.
 
\section{Conclusions}

Our motivation for this work is to use Bianchi universes to search for anisotropies in the Lema{\^i}tre-Hubble diagram of type 1a supernovae. These universes are necessarily close to the maximally symmetric Friedman universes and, if we want to add only the minimal number of degrees of freedom to the fit of the standard model, the Bianchi universes must have four isometries; hence our definition of minimal symmetry breaking of the cosmological principle. 
The other extreme are maximal symmetry breakings. They break all symmetries and may have an infinite number of degrees of freedom,    `geographic deviations'.

As an aside we remark that the criterium of a minimal number of broken symmetries can also be tried on the standard model of particle physics. There however it fails. 
Indeed, the hyper-charged isospin-doublet color-singlet of scalars breaks the (3+1+8\ =\ 12)-dimensional gauge group 
\bb
SU(2)_\text{isospin}\times U(1)_\text{hyper-charge}\times SU(3)_\text{color}
\ee
 spontaneously down to the (1+8\ =\ 9)-dimensional little group
\bb
 U(1)_\text{electric charge}\times SU(3)_\text{color}.
\ee
This symmetry breaking is {\it not} minimal and the minimal breaking is induced by a hyper-charged isospin- and color-singlet with (3+8\ =\ 11)-dimensional little group
\bb
SU(2)_\text{isospin}\times SU(3)_\text{color}.
\ee

By Luigi Bianchi's beautiful classification of Riemannian 3-spaces admitting  isometry groups \cite{Bianchi}, the minimal symmetry breaking of the cosmological principle only leaves the axial Bianchi I, V and IX universes, see the cartoon in figure \ref{cart}.
\renewcommand{\thefigure}{3}
\begin{figure}[htbp]\hspace{17mm}
\includegraphics[width=3.3cm, height=2.1cm]{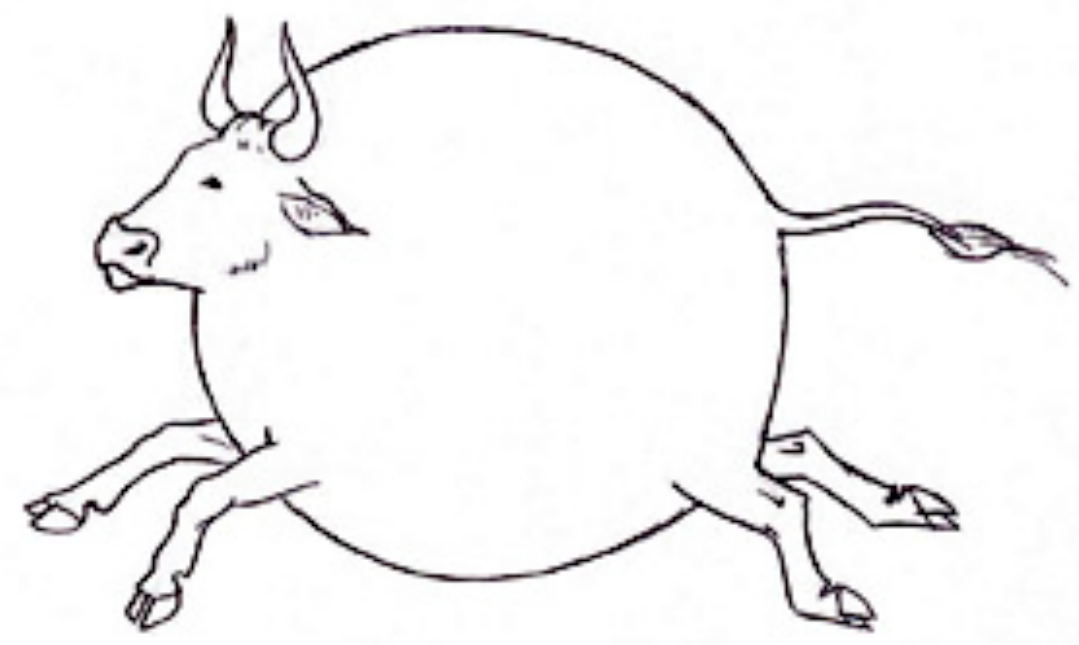}\hspace{9mm}
\includegraphics[width=4.2cm, height=2.cm]{Cow.pdf}\hspace{14mm}
\includegraphics[width=3.3cm, height=2.0cm]{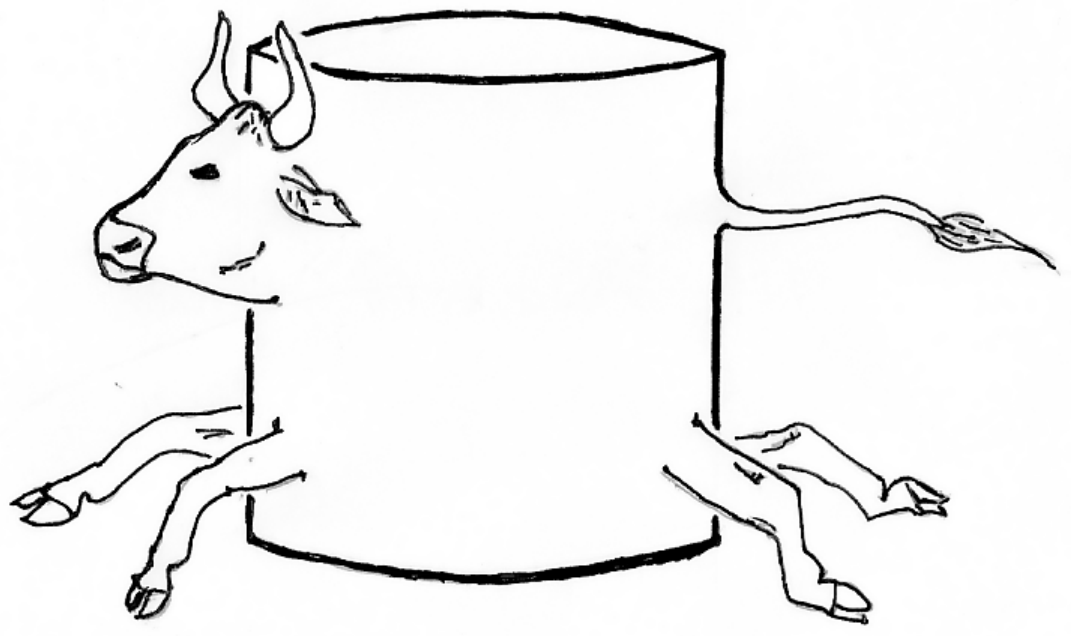}
\caption[]{Robertson-Walker;   \hspace{6mm}
axial Bianchi I, V, IX;  \hspace{6mm}
Bianchi II, III, IV, VI$_h$,\,VIII
}
 \label{cart}
\end{figure}

Axial Bianchi V universes are incompatible with comoving dust, an asymmetry between negative and positive curvature that we find most remarkable.

Consequently we computed only  the Lema{\^i}tre-Hubble diagram for  axial Bianchi IX universes. In the limit of zero curvature, it reproduces the diagram for axial Bianchi I universes.

We completely agree with Aluri, Cea et al.
\cite{cea22} and join their plea for tests of the cosmological principle in synergy of all available cosmological probes.  Axial Bianchi IX universes must be among the first choices of such a test.

On the theoretical side we remain intrigued by a minimal symmetry breaking of maximally symmetric spacetimes starting with the generalization of Fubini's theorem \cite{Fubini} to pseudo-Riemannian manifolds. This generalization has been achieved in 2003 by G. S. Hall \cite{hall}.

\vspace{3mm}
\noindent
{\bf Acknowledgements:} It is a pleasure to thank Yves Cornulier for his friendly advice via MathOverflow and Christoph Stephan for the reference to Hall's work \cite{hall}. We are also grateful to  Eoin \'O Colg\'ain for his constructive comments. Finally we acknowledge discouraging comments by
 an anonymous referee, c.f. appendix.

The project leading to this publication has received funding from Excellence Initiative of Aix Marseille University - A*MIDEX, a French "Investissements d'Avenir" programme (AMX-19-IET-008 - IPhU).

\vspace{2mm}
\noindent
{\bf Data availability:} The data underlying this article will be shared on reasonable request to the corresponding author.

%\small
\footnotesize
\appendix
 \section{Appendix}
 
 \setcounter{equation}{0}
 
 \noindent
$\bullet$ 23 November 2022

submission to General Relativity and Gravitation

\vspace{5mm}
\noindent
$\bullet$ 24 March 2023

Dear Dr Sch\"ucker,

Your manuscript entitled ``Axial Bianchi IX and its Lemaitre-Hubble diagram'' has now been assessed. 
Regrettably, the above submission has been rejected for publication in General Relativity and Gravitation.

Thank you for the opportunity to consider your work. I am sorry that we cannot be more positive on this occasion and hope you will not be deterred from submitting future work to General Relativity and Gravitation.

Kind regards,

Mairi Sakellariadou

Editor, 
General Relativity and Gravitation

\vspace{3mm}
Reviewer Comments:

The authors consider Bianchi models, notably Bianchi type V and the locally rotationally symmetric (LRS) Bianchi type IX models, with a focus on the latter, with dust and a cosmological constant as the matter source. They investigate the Einsten field equations as well as various cosmographic aspects. Unfortunately, they seem to be completely unaware that this is an old well-developed field with a vast literature (they have a significant number of references, but most relevant ones are missing). For this reason, the paper is unacceptable for publication (in addition, there are numerous erroneous statements in the paper that would have been avoided if the authors had been aware of the literature). The authors simply need to educate themselves in the field before attempting to produce a contribution to it. Here are some of the references in the field the authors need to use as a starting point for a literature search in order for to familiarize themselves with the field. 
It is well known that the diagonal Bianchi type V models with an orthogonal (non-tilted) fluid with a barotropic equation of state, including a cosmological constant, are solvable. For how to do this as well as how to find all explicit solutions in spatially homogeneous cosmology, see the paper ``Exact hypersurface-homogeneous solutions in cosmology and astrophysics'' by C. Uggla et al., Physical Review D 51 (10), 5522. For a complete description of the solution space of the LRS type IX models, see ``Compactified and reduced dynamics for locally rotationally symmetric Bianchi type IX perfect fluid models'' by Uggla and Zur-Muhlen, Classical and Quantum Gravity 7 (8), 1365 and, especially, ``Matter and dynamics in closed cosmologies'' by Heinzle, [Uggla] et al., Physical Review D 71 (8), 083506. To do cosmography in spatially homogeneous cosmology there is no need to introduce coordinate dependent one-forms, for an example of how to do this geometrically, see ``A dynamical systems approach to geodesics in Bianchi cosmologies'' by Nilsson, [Uggla] et al., General Relativity and Gravitation 32, 1319-1343 and ``An almost isotropic cosmic microwave temperature does not imply an almost isotropic universe'' by Nilsson, [Uggla]  et al., The Astrophysical Journal Letters 522, L1. For a standard review of Bianchi cosmology, see the book ``Dynamical systems in cosmology'' Eds. Wainwright and Ellis, with many additional relevant references (and since it is well cited this reference can also be used to find more recent references). Furthermore, the models the authors consider are special cases of more general matter models studied by e.g. A. Coley and S. Hervik. The above should provide a starting point for the authors to make themselves familiar with the field, if they choose to pursue it.

\vspace{5mm}
\noindent
$\bullet$ 28 March 2023

Dear Professeure Sakellariadou,

thank you for your referee's report. We regret that after three and a half months we only obtained a report that can be written in half an hour. 

Indeed it is said that our paper is unacceptable for publication because it misses the most relevant references of an old well-developed field. Your referee is kind enough to give us a precise list of those, mostly by Uggla and collaborators. We are well aware of and appreciate this field: the dynamical systems approach. However, to the best of our knowledge, this approach has never been used to compute any Lema\^itre-Hubble diagram, certainly not in the suggested papers by Uggla and collaborators.

In addition, it is said, there are numerous erroneous statements in our paper. We would be eager to correct these statements. But not a single one is given in the report.

Your referee invites us to read the book ``Dynamical systems in cosmology''. We own a copy of it and use it since years.

As stated explicitly in the title and in the abstract of our paper, we compute the Lema\^itre-Hubble diagrams for the axial Bianchi IX universes. These diagrams are ready to be confronted with the supernova data to be obtained by the Vera Rubin Observatory starting in two or three years.

We maintain that the dynamical systems approach and in particular the list of suggested papers by Uggla and collaborators are of no help in computing the Lema\^itre-Hubble diagrams for the axial Bianchi IX universes. Your referee disagrees and we would like him 

-1- to point out concrete equations in his list of references that could be of any help in simplifying our computations of the Lema\^itre-Hubble diagrams for the axial Bianchi IX universes;

-2- to list the numerous erroneous statements in our paper;

-3- to point out where our solution of Bianchi V universes with cosmological constant and dust can be found in ``Dynamical systems in cosmology''.

Without these three items, we feel that his report it is empty.

We hope that you share our feelings and that you can convince your referee to deliver the three items.

Sincerely yours,

 Galliano Valent, Andr\'e Tilquin and Thomas Sch\"ucker
 
\vspace{5mm}
\noindent
$\bullet$ 25 April 2023

Dear Dr Sch\"ucker,

Please find attached document. We remain with our initial decision.

Best wishes,

M Sakellariadou, 
EiC

\vspace{3mm}

The authors of the paper ``Axial Bianchi IX and its Lema\^itre-Hubble diagram'' asked me as the referee to complement my referee report with some specifics. Let me first say that when I was asked to referee the paper I did so the next day and it took all day to read the paper and comment on it and I then sent in the referee report promptly.

I understand that the authors spent a lot of time on their paper and that the referee report must have come as a disappointment. Moreover, their paper is better than many published papers, but they should not in my opinion have been published either, and I stand by my assessment that their paper is not suitable for publication.

Let me comment on their points in reverse order and thereby begin with their point 3, about the Bianchi type V solution. As stated in my report, it is well known that these models are explicitly solvable for a cosmological constant and an orthogonal (non-tilted) perfect fluid with an arbitrary barotropic equation of state $p=p(\rho )$, where dust is just a special case, as shown implicitly in the paper ``Exact hypersurface-homogeneous solutions in cosmology and astrophysics'' by C. Uggla et al., Physical Review D 51 (10), 5522 (UJR), not the Dynamical Systems in Cosmology book. Moreover, this is obvious already before doing any calculations. The intrinsic curvature of Bianchi type V is isotropic (as illustrated by the authors themselves after eq. 8 since the spatial curvature $\propto {\delta^\alpha }_\beta $)  on each spatial slice. Non-tilted dust, like any barotropic fluid, also yields an isotropic contribution. It is for these reasons the Hamiltonian in UJR only involves the scale factor and that anisotropy is given by the extrinsic curvature. This makes it obvious that it is possible to solve this case by integrating a single first order differential equation. Let me use UJR to sketch the derivation of the solution, since UJR provides instructions for how to obtain all known exact solutions in spatially homogeneous cosmology in as simple form as possible, where type V is just a simple special case, as pointed out in the Exact Solution book by MaCallum et al.

We hence follow section II in UJR. For simplicity let us use the Taub gauge with the lapse $N=N_T=12e^{3\beta ^0}$ where $e^{\beta ^0}$  is the cosmological scale factor (this time gauge is, modulo an unimportant constant factor, what the authors use). This leads to the following Bianchi type V line-element
\bb
ds^2=-N_T^2dt_T^2+e^{2\beta ^0}\lb e^{2\sqrt{3}\beta ^\times}\omega _1^2+e^{-2\sqrt{3}\beta ^\times}\omega _2^2+\omega _3^2\rb,
\eee
where $e^{\beta ^0}$   thereby is the cosmological scale factor ($\beta ^\times$ is sometimes referred to as $\beta ^-$ in the literature that uses Misner metric parametrizations), and the Hamiltonian
\bb
H_T=\,\frac{1}{2}\lb  -p_0^2+p_\times^2+V_T(\beta ^0)\rb=0,
\eee
 where
 \bb
 V_T=12^2e^{4\beta ^0}+48\kappa e^{6\beta ^0}\rho (\beta ^0)+48e^{6\beta ^0}\Lambda ,
 \eee
where $12^2e^{4\beta ^0}$ comes from the spatial curvature and $\rho $ is the energy density of the fluid.
For a barotropic equation of state $p=p(\rho )$ local energy conservation yields
\bb
\,\frac{d\rho }{d\beta ^0} =-3\rho (\rho +p(\rho ))
\eee
which for a linear equation of state
$p=(\gamma -1 )\rho $
  leads to 
  $\rho =\rho _0e^{-3\gamma \beta ^0}$,
   where dust $\gamma =1$ then yields
   $\rho = \rho _0e^{-3\beta ^0}$.
 Since $\beta ^\times$  is a cyclic variable, the conjugate momentum $p_\times$ = constant. It then follows that
 $\beta ^\times = p_\times t_T$, since 
 $d\beta ^\times /dt_T=p_\times$
 due to the Hamiltonian equations, which together with the Hamiltonian constraint (2) yields the solution in the form of the authors, where their
 $K\propto p_\times$.
 
This, however, is not the best way to present the solution with respect to the claimed purpose of the authors of studying things like the redshift. Note that $\beta ^0$  leads to that $e^{-\beta ^0}=1+z$  for the FLRW isotropic models (set the scale factor 
$R=e^{\beta ^0}=1$,
 and hence
 $\beta ^0=0$,
 at the present time, as is usually done). This strongly suggests that, with the authors interests in mind, one should use the e-fold time $\beta ^0$ as the time variable. As discussed in section III B in UJR this is an intrinsic gauge, which was used in ref. [22] in UJR to present the type V solution for an orthogonal perfect fluid with a linear equation of state. Just add the cosmological constant to that solution and specialize to dust and you are done. However, since the authors wanted details, let me show how to get the solution explicitly.
 
First introduce
\bb
F:=\frac{1}{p_\times^2+V_T(\beta ^0)}.
 \eee
 Then note that
 $p_0=-d\beta ^0/dt_T$ and hence that, as follows from (2),
 \bb
 \frac{dt_T}{d\beta ^0}=\sqrt{F(\beta ^0)},
 \eee
 which leads to
 \bb
 \beta ^\times(\beta ^0)=p_\times\int\sqrt{F(\tilde\beta ^0)}d\tilde\beta ^0
 \eee
 and the line element
 \bb
 ds^2=-144e^{6\beta ^0}F(\beta ^0)(d\beta ^0)^2
 +e^{2\beta ^0}\lb e^{2\sqrt{3}\beta ^\times}\omega _1^2+e^{-2\sqrt{3}\beta ^\times}\omega _2^2+\omega _3^2\rb,
\eee
followed by specializing $\rho $ in $F(\beta ^0)$ to dust.

Let me next comment and elaborate on their point 2 about erroneous statements. That there are errors is exemplified by the authors' comments in their section 2 about the parameter $h$ for the class B Bianchi type $V I_h$ and $V II_h$ models; $h$ is a group invariant that is negative for type $V I_h$ and positive for $V II_h$. Arguably  
more annoying than errors, however, are many well known statements that are presented in a way as giving an impression that they are new. For example, in the abstract(!) ``Remarkably, negative curvatures are excluded by this minimal symmetry breaking in presence of comoving dust.'' This has been known since almost the dawn of spatially homogeneous cosmology. Moreover, just compare this statement with the content in ``Irrotational type V models'', by C. G. Hewitt and J. Wainwright in Phys. Rev. D 46, 4242, which is easy to generalize to also include a cosmological constant.

Then take the bullet points in the introduction; as regards the first point, lots of work has been done on LRS models, and as regards point 2, the Hubble-normalized dynamical systems approach yields the $\Lambda $CDM model as a straight line on the boundary of all spatially homogeneous models, which is easily perturbed to straightforwardly yield perturbed Lema\^itre-Hubble diagrams for all spatially homogenous models. This brings me to another even more annoying thing. The authors are not doing what they claim to do in the title and abstract. What they actually do is just a linear perturbation in certain variables to obtain expressions for the redshift and the apparent luminocity [luminosity]. This motivated my reference to ``Matter and dynamics in closed cosmologies'' by Heinzle, [Uggla] et al., Physical Review D 71 (8), 083506. There the entire solution space of the LRS type IX model (dust and $\Lambda $ is just a special case) is rigorously described and shown to exhibit a remarkably complex structure (which shows that a complete redhift and apparent luminocity description of these models constitutes a formidable task), but the closed FLRW solution is just a simple 2-dimensional subset in the formulation in this paper; this subset is easily perturbed and thereby, in contrast to the authors work, situates the perturbation and associated observables, such as Lema\^itre- Hubble diagrams, in the full state space of these models, which makes a clear interpretation of the results possible.

Worst of all, most of the material in the paper is irrelevant for the stated purpose of the paper, and is, moreover, well known. To illustrate this, let me address point 1 (although some of the above already addresses that point). {\it Observables are gauge invariant.} Hence there is no need for {\it explicit coordinates} for 1-forms etc. that takes up much of the space of the paper. Thus my comment ``To do cosmography in spatially homogeneous cosmology there is no need to introduce coordinate dependent one-forms, for an example of how to do this geometrically, see ``A dynamical systems approach to geodesics in Bianchi cosmologies'' by Nilsson, [Uggla]  et al., General Relativity and Gravitation 32, 1319-1343 and ``An almost isotropic cosmic microwave temperature does not imply an almost isotropic universe'' by Nilsson, [Uggla]  et al., The Astrophysical Journal Letters 522, L1.'' These papers illustrate explicitly that there is no need to introduce coordinates to obtain results for observables (as an application the focus in those papers was on temperature distributions, which is actually more challenging than redshifts and apparent luminocities, which constitute straightforward additional illustrations
--- thus these papers directly answers the authors point 1), especially since coordinates just obfuscate what is going on observationally.

Finally, I find the motivation for what the authors actually do somewhat lacking. Nevertheless, I am actually rather sympathetic to what seems to be the authors goal in a broader sense, namely connecting spatially homogenous cosmology to cosmography. However, this is again a topic with a history. Notably it was pioneered by Ellis, MacCallum, Barrow during the 1970s, so at the very least that would have to be a starting point for such a project (again, the authors has none of the relevant references). Since present observations indicate that the Universe is accelerating, where a cosmological constant is the simplest explanation, it might be worthwhile to revisit this topic again using up to date observations, but one should then do so with modern techniques and with the benefits of the present status and knowledge about the field of spatially homogeneous cosmology, as indicated by my above comments. Even though it might be frustrating, I urge the authors to take a step back and take my referee report and the above seriously, since this would allow them to do much better than the present submitted paper.

\vspace{5mm}
\noindent
$\bullet$ 1 May 2023

Dear Professeure Sakellariadou,

thank your for the reply of your referee. 

In his initial report he claims, ``there are numerous erroneous statements in the paper'' and we asked him to point them out. In his reply he claims three errors:

\{1\}:  ``Let me comment and elaborate on their erroneous statements. That there are errors is exemplified by the authors' comments in their section 2 about the parameter $h$ for the class B Bianchi type $V I_h$ and $V II_h$ models; $h$ is a group invariant that is negative for type $V I_h$ and positive for $V II_h$.''

In the literature there are several conventions for the parameter $h$. We are sorry that we have not chosen the one preferred by your referee. But we clearly state our source, [61] Terzis (2013), and we make our conventions explicit in table (5). 

\{2\}:  ``Even more annoying than errors the statement `Remarkably, negative curvatures are excluded by this minimal symmetry breaking in presence of comoving dust.' presented in a way as giving an impression that it is new.''

We did not claim that it is new, but give credit to [57] Farnsworth (1967) for it. And yes we do think that this fact is remarkable, in particular in the light of our reference [60] Akarsu, Di Valentino, Kumar, Ozyigit \& Sharma (2021), which describes a Bianchi V universe with comoving dust.

\{3\}:  ``This brings me to another even more annoying thing. The authors are not doing what they claim to do in the title and abstract. What they actually do is just a linear perturbation in certain variables to obtain expressions for the redshift and the apparent luminocity.''

Our equations (76) and (94) give the redshift and the apparent luminosity before linearization of the scale parameters. 

The ensuing Lema\^itre-Hubble diagram for the axial Bianchi IX model is the main result of our paper and, to the best of our knowledge, it is new.

\vspace{3mm}
Where are the numerous erroneous statements in our paper?

\vspace{3mm}
To put our result into perspective we recall a controversy concerning the apparent luminosity in Bianchi I universes. It was calculated by [48] Saunders (1969) and by [50]  Koivisto \& Mota (2008) who found a different result. Our ab initio computation, [47] Sch\"ucker, Tilquin \& Valent (2014) confirmed Saunders' result and [52] Fleury, Pitrou \& Uzan (2015) confirmed our result by another calculation.

We also reject your referee's critique of missing motivation:

``Finally, I find the motivation for what the authors actually do somewhat lacking.''

We give our experimental motivation in the introduction and with more details at the end of our section 9: In a few years new instruments like the Vera Rubin Observatory and the James Webb Space Telescope will measure the Lema\^itre-Hubble diagram with unprecedented precision and we feel encouraged to propose cosmological models that can be tested by the new data.

We give a theoretical motivation in the introduction and with more details in our section 2 by defining a minimal symmetry breaking of the cosmological principle, which in turn we motivate with a 2-dimensional example, the shape of our planet's surface. 

In his reply your referee reiterates his advice that we read and cite the historic papers:

``The authors simply need to educate themselves in the field before attempting to produce a contribution to it.''

``... a topic with a history. Notably it was pioneered by Ellis, MacCallum, Barrow during the 1970s, so at the very least that would have to be a starting point for such a project (again, the authors has none of the relevant references).''

In our paper we cite:
[1] Ellis \& MacCallum (1969),
[2] Hawking (1969), 
[3] Collins \& Hawking (1973),
[4] Barrow, Juszkiewicz \& Sonoda (1985),
[5] Barrow \& Levin (1997),
[21] Lema\^itre (1933),
[48] Saunders (1969),
[61] Farnsworth (1967),
[62] King \& Ellis (1973). 

Does he really claim to know that we have not read these? 

In his reply your referee makes remarks about his personal tastes:

``there is no need to introduce coordinates to obtain results for observables''

``especially since coordinates just obfuscate what is going on observationally''\\
and he presents his alternative to our calculation of the solution to Einstein's equations in Bianchi V universes. His calculation is two pages long and builds on results in a paper by Uggla et al. (1995), that you must have read before understanding his two pages. Our calculation is one page long, pedestrian and ab initio.

There are only correct and false calculations and this is the first time in our career that we must defend our personal choices concerning proofs:

Our paper proposes a test of axial Bianchi IX universes in a form ready to be used by experimentalists and we wanted our calculations to be as self-contained and as user-friendly to experimentalists as possible. (One of us is experimentalist.) For the sake of communication, we stand by our choices: low tech.

Unfortunately our exchange with your referee is becoming sterile and we kindly ask you, more than four months after submission, for the opinion of a second referee on our paper and on this exchange with your first referee.

Sincerely yours, Galliano Valent, Andr\'e Tilquin and Thomas Sch\"ucker

\vspace{5mm}
\noindent
$\bullet$ No reply by Prof. Sakellariadou


\begin{thebibliography}{10}

\bibitem{EMC}
G. Ellis and M. A. MacCallum,
``A class of homogeneous cosmological models,''
 Comm. Math. Phys. {\bf 12} (1969) 108.
 
\bibitem{hawk}
S. Hawking,
 ``On the rotation of the Universe,''
  Mon.\ Not.\ Roy.\ Astron.\ Soc.\  {\bf 142} (1969) 129
  doi:10.1093/mnras/142.2.129
 
 \bibitem{ch}
C.~B. Collins and S. Hawking,
 ``The rotation and distortion of the Universe,''
  Mon.\ Not.\ Roy.\ Astron.\ Soc.\  {\bf 162} (1973) 305
  doi:10.1093/mnras/162.4.307

 \bibitem{bjs}
J.~D. Barrow, R. Juszkiewicz and D.~H. Sonoda,
 ``Universal rotation - How large can it be?''
  Mon.\ Not.\ Roy.\ Astron.\ Soc.\  {\bf 213} (1985) 917
  doi:10.1093/mnras/213.4.917
 
  \bibitem{Barrow:1997vu}
J.~D.~Barrow and J.~J.~Levin,
``Geodesics in open universes,''
Phys. Lett. A \textbf{233} (1997) 169
doi:10.1016/S0375-9601(97)00504-5
[arXiv:astro-ph/9704041 [astro-ph]].

\bibitem{Jaffe}
T.~R.~Jaffe, S.~Hervik, A.~J.~Banday and K.~M.~Gorski,
``On the viability of Bianchi type viih models with dark energy,''
Astrophys. J. \textbf{644} (2006) 701
doi:10.1086/503893
[arXiv:astro-ph/0512433 [astro-ph]].

  \bibitem{Bridges}
M.~Bridges, J.~D.~McEwen, A.~N.~Lasenby and M.~P.~Hobson,
``Markov chain Monte Carlo analysis of Bianchi VII(h) models,''
Mon. Not. Roy. Astron. Soc. \textbf{377} (2007) 1473
doi:10.1111/j.1365-2966.2007.11616.x
[arXiv:astro-ph/0605325 [astro-ph]].

\bibitem{Pontzen:2007ii}
A.~Pontzen and A.~Challinor,
``Bianchi Model CMB Polarization and its Implications for CMB Anomalies,''
Mon. Not. Roy. Astron. Soc. \textbf{380} (2007) 1387
doi:10.1111/j.1365-2966.2007.12221.x
[arXiv:0706.2075 [astro-ph]].

\bibitem{CampanelliCMB}
L.~Campanelli, P.~Cea and L.~Tedesco,
``Cosmic Microwave Background Quadrupole and Ellipsoidal Universe,''
Phys. Rev. D \textbf{76} (2007) 063007
doi:10.1103/PhysRevD.76.063007
[arXiv:0706.3802 [astro-ph]].

\bibitem{ppuCMB}
T.~S.~Pereira, C.~Pitrou and J.~P.~Uzan,
``Theory of cosmological perturbations in an anisotropic universe,''
JCAP \textbf{09} (2007) 006
doi:10.1088/1475-7516/2007/09/006
[arXiv:0707.0736 [astro-ph]].

\bibitem{Pontzen:2010eg}
A.~Pontzen and A.~Challinor,
``Linearization of homogeneous, nearly-isotropic cosmological models,''
Class. Quant. Grav. \textbf{28} (2011) 185007
doi:10.1088/0264-9381/28/18/185007
[arXiv:1009.3935 [gr-qc]].

\bibitem{planck}
P.~A.~R.~Ade \textit{et al.} [Planck],
``Planck 2013 results. XXVI. Background geometry and topology of the Universe,''
Astron. Astrophys. \textbf{571} (2014) A26
doi:10.1051/0004-6361/201321546
[arXiv:1303.5086 [astro-ph.CO]].

\bibitem{Russell}
E.~Russell, C.~B.~ {K{\i}l{\i}n{\c c}} and O.~K.~Pashaev,
``Bianchi I model: an alternative way to model the present-day Universe,''
Mon. Not. Roy. Astron. Soc. \textbf{442} (2014) no.3 2331
doi:10.1093/mnras/stu932
[arXiv:1312.3502 [astro-ph.CO]].

  \bibitem{cea1}
  P.~Cea,
``The Ellipsoidal Universe in the Planck Satellite Era,''
Mon. Not. Roy. Astron. Soc. \textbf{441} (2014) no.2 1646
doi:10.1093/mnras/stu687
[arXiv:1401.5627 [astro-ph.CO]].

\bibitem{wmap}
D.~Saadeh, S.~M.~Feeney, A.~Pontzen, H.~V.~Peiris and J.~D.~McEwen,
``A framework for testing isotropy with the cosmic microwave background,''
Mon. Not. Roy. Astron. Soc. \textbf{462} (2016) no.2 1802
doi:10.1093/mnras/stw1731
[arXiv:1604.01024 [astro-ph.CO]].

  \bibitem{cea2}
  P.~Cea,
``Confronting the Ellipsoidal Universe to the Planck 2018 Data,''
Eur. Phys. J. Plus \textbf{135} (2020) no.2, 150
doi:10.1140/epjp/s13360-020-00166-5
[arXiv:1909.05111 [astro-ph.CO]].

\bibitem{cea3}
P.~Cea,
``The Ellipsoidal Universe and the Hubble tension,''
[arXiv:2201.04548 [astro-ph.CO]].

\bibitem{cea4}
P.~Cea,
``CMB two-point angular correlation function in the Ellipsoidal Universe,''
[arXiv:2203.14229 [astro-ph.CO]].

\bibitem{pp}
T.~S.~Pereira and C.~Pitrou,
``Bianchi spacetimes as supercurvature modes around isotropic cosmologies,''
Phys. Rev. D \textbf{100} (2019) no.12, 123534
doi:10.1103/PhysRevD.100.123534
[arXiv:1909.13688 [gr-qc]].

\bibitem{Bianchi}
Luigi Bianchi, ``Sugli spazi a tre dimensioni che ammettono un gruppo continuo di movimenti,'' Memorie di Matematica e di Fisica della Societa Italiana delle Scienze {\bf 11} (1898) 267352;\\
English translation by R.T. Jantzen,``On the three-dimensional spaces which admit a continuous group of motions,'' Gen. Rel. Grav. {\bf 33} (2001) 2171, {\tt http://www34.homepage.villanova.edu/robert.jantzen/bianchi/}
  
   \bibitem{lemaitre}
George Lema\^itre,
``L'univers en expansion,''
Annales Soc. Sci. Bruxelles A \textbf{53} (1933), 51-85
doi:10.1023/A:1018855621348;\\
English translation by M. A. H. MacCallum,``The expanding universe,'' Gen. Rel. Grav. \textbf{29} (1997) 641.

  \bibitem{cospar}
 C.~Quercellini, M.~Quartin and L.~Amendola,
  ``Possibility of Detecting Anisotropic Expansion of the Universe by Very Accurate Astrometry Measurements,''
  Phys.\ Rev.\ Lett.\  {\bf 102} (2009) 151302
  [arXiv:0809.3675 [astro-ph]].
  
  \bibitem{font}
  M.~Fontanini, E.~J.~West and M.~Trodden,
  ``Can Cosmic Parallax Distinguish Between Anisotropic Cosmologies?,''
  Phys.\ Rev.\ D {\bf 80} (2009) 123515
  [arXiv:0905.3727 [astro-ph.CO]].
  
  \bibitem{tedesco}
   L.~Campanelli, P.~Cea, G.~L.~Fogli and L.~Tedesco,
  ``Cosmic Parallax in Ellipsoidal Universe,''
  Mod.\ Phys.\ Lett.\ A {\bf 26} (2011) 1169
  [arXiv:1103.6175 [astro-ph.CO]].
  
   \bibitem{dar}
   J.~Darling,
``The Hubble Expansion is Isotropic in the Epoch of Dark Energy,''
Mon. Not. Roy. Astron. Soc. \textbf{442} (2014) L66
doi:10.1093/mnrasl/slu057
[arXiv:1404.3735 [astro-ph.CO]].

\bibitem{mppuDrift}
O.~H.~Marcori, C.~Pitrou, J.~P.~Uzan and T.~S.~Pereira,
``Direction and redshift drifts for general observers and their applications in cosmology,''
Phys. Rev. D \textbf{98} (2018) no.2, 023517
doi:10.1103/PhysRevD.98.023517
[arXiv:1805.12121 [astro-ph.CO]].


\bibitem{hemi1}
T.~S.~Kolatt and O.~Lahav,
 ``Constraints on cosmological anisotropy out to z=1 from supernovae ia,''
  Mon.\ Not.\ Roy.\ Astron.\ Soc.\  {\bf 323} (2001) 859
  [astro-ph/0008041].
  
  \bibitem{hemi2}
  D.~J.~Schwarz and B.~Weinhorst,
  ``(An)isotropy of the Hubble diagram: Comparing hemispheres,''
  Astron.\ Astrophys.\  {\bf 474} (2007) 717
  [arXiv:0706.0165 [astro-ph]].
  
  \bibitem{hemi3}
  I.~Antoniou and L.~Perivolaropoulos,
  ``Searching for a Cosmological Preferred Axis: Union2 Data Analysis and Comparison with Other Probes,''
  JCAP {\bf 1012} (2010) 012
  [arXiv:1007.4347 [astro-ph.CO]].
  
  \bibitem{hemi4}
  B.~Kalus, D.~J.~Schwarz, M.~Seikel and A.~Wiegand,
 ``Constraints on an\-iso\-tropic cosmic expansion from supernovae,''
  Astron.\ Astrophys.\  {\bf 553} (2013) A56
  [arXiv:1212.3691 [astro-ph.CO]].
  
  \bibitem{hemi5} X.~Yang, F.~Y.~Wang and Z.~Chu,
``Searching for a preferred direction with Union2.1 data,''
Mon. Not. Roy. Astron. Soc. \textbf{437} (2014) no.2, 1840
doi:10.1093/mnras/stt2015
[arXiv:1310.5211 [astro-ph.CO]].

  \bibitem{hemi6}
  J.~Beltran Jimenez, V.~Salzano and R.~Lazkoz,
``Anisotropic expansion and SNIa: an open issue,''
Phys. Lett. B \textbf{741} (2015) 168
doi:10.1016/j.physletb.2014.12.031
[arXiv:1402.1760 [astro-ph.CO]].

\bibitem{java}
B.~Javanmardi, C.~Porciani, P.~Kroupa and J.~Pflamm-Altenburg,
``Probing the isotropy of cosmic acceleration traced by Type Ia supernovae,''
Astrophys. J. \textbf{810} (2015) no.1, 47
doi:10.1088/0004-637X/810/1/47
[arXiv:1507.07560 [astro-ph.CO]].

\bibitem{Krish}
C.~Krishnan, R.~Mohayaee, E.~\'O.~Colg\'ain, M.~M.~Sheikh-Jabbari and L.~Yin,
``Hints of FLRW breakdown from supernovae,''
Phys. Rev. D \textbf{105} (2022) no.6, 063514
doi:10.1103/PhysRevD.105.063514
[arXiv:2106.02532 [astro-ph.CO]].

\bibitem{Secr}
N.~J.~Secrest, S.~von Hausegger, M.~Rameez, R.~Mohayaee, S.~Sarkar and J.~Colin,
``A Test of the Cosmological Principle with Quasars,''
Astrophys. J. Lett. \textbf{908} (2021) no.2, L51
doi:10.3847/2041-8213/abdd40
[arXiv:2009.14826 [astro-ph.CO]].

\bibitem{Luon}
O.~Luongo, M.~Muccino, E.~\'O.~Colg\'ain, M.~M.~Sheikh-Jabbari and L.~Yin,
``Larger H0 values in the CMB dipole direction,''
Phys. Rev. D \textbf{105} (2022) no.10, 103510
doi:10.1103/PhysRevD.105.103510
[arXiv:2108.13228 [astro-ph.CO]].

  \bibitem{ppuWeakL}
C.~Pitrou, T.~S.~Pereira and J.~P.~Uzan,
``Weak-lensing by the large scale structure in a spatially anisotropic universe: theory and predictions,''
Phys. Rev. D \textbf{92} (2015) no.2, 023501
doi:10.1103/PhysRevD.92.023501
[arXiv:1503.01125 [astro-ph.CO]].

\bibitem{Mig}
K.~Migkas, G.~Schellenberger, T.~H.~Reiprich, F.~Pacaud, M.~E.~Ramos-Ceja and L.~Lovisari,
``Probing cosmic isotropy with a new X-ray galaxy cluster sample through the $L_{\text{X}}-T$ scaling relation,''
Astron. Astrophys. \textbf{636} (2020) A15
doi:10.1051/0004-6361/201936602
[arXiv:2004.03305 [astro-ph.CO]].

\bibitem{Mig2}
K.~Migkas, F.~Pacaud, G.~Schellenberger, J.~Erler, N.~T.~Nguyen-Dang, T.~H.~Reiprich, M.~E.~Ramos-Ceja and L.~Lovisari,
``Cosmological implications of the anisotropy of ten galaxy cluster scaling relations,''
Astron. Astrophys. \textbf{649} (2021), A151
doi:10.1051/0004-6361/202140296
[arXiv:2103.13904 [astro-ph.CO]].

\bibitem{siew}
T.~M.~Siewert, M.~Schmidt-Rubart and D.~J.~Schwarz,
``Cosmic radio dipole: Estimators and frequency dependence,''
Astron. Astrophys. \textbf{653} (2021), A9
doi:10.1051/0004-6361/202039840
[arXiv:2010.08366 [astro-ph.CO]].

\bibitem{cup}
G.~Cusin, C.~Pitrou and J.~P.~Uzan,
``Anisotropy of the astrophysical gravitational wave background: Analytic expression of the angular power spectrum and correlation with cosmological observations,''
Phys. Rev. D \textbf{96} (2017) no.10, 103019
doi:10.1103/PhysRevD.96.103019
[arXiv:1704.06184 [astro-ph.CO]].

\bibitem{Cusin:2018rsq}
G.~Cusin, I.~Dvorkin, C.~Pitrou and J.~P.~Uzan,
``First predictions of the angular power spectrum of the astrophysical gravitational wave background,''
Phys. Rev. Lett. \textbf{120} (2018) 231101
doi:10.1103/PhysRevLett.120.231101
[arXiv:1803.03236 [astro-ph.CO]].

\bibitem{Cusin:2019jhg}
G.~Cusin, I.~Dvorkin, C.~Pitrou and J.~P.~Uzan,
``Stochastic gravitational wave background anisotropies in the mHz band: astrophysical dependencies,''
Mon. Not. Roy. Astron. Soc. \textbf{493} (2020) no.1, L1
doi:10.1093/mnrasl/slz182
[arXiv:1904.07757 [astro-ph.CO]].

\bibitem{Cusin:2019jpv}
G.~Cusin, I.~Dvorkin, C.~Pitrou and J.~P.~Uzan,
``Properties of the stochastic astrophysical gravitational wave background: astrophysical sources dependencies,''
Phys. Rev. D \textbf{100} (2019) no.6, 063004
doi:10.1103/PhysRevD.100.063004
[arXiv:1904.07797 [astro-ph.CO]].

\bibitem{pcu}
C.~Pitrou, G.~Cusin and J.~P.~Uzan,
``Unified view of anisotropies in the astrophysical gravitational-wave background,''
Phys. Rev. D \textbf{101} (2020) no.8, 081301
doi:10.1103/PhysRevD.101.081301
[arXiv:1910.04645 [astro-ph.CO]].

\bibitem{Colgain}
E.~\'O.~Colg\'ain,
``Probing the Anisotropic Universe with Gravitational Waves,''
[arXiv:2203.03956 [astro-ph.CO]].

  \bibitem{stv}
T.~Sch\"ucker, A.~Tilquin and G.~Valent,
``Bianchi~I meets the Hubble diagram,''
Mon. Not. Roy. Astron. Soc. \textbf{444} (2014) 2820
doi:10.1093/mnras/stu1656
[arXiv:1405.6523 [astro-ph.CO]].

\bibitem{Sa} P. T. Saunders, ``Observations in some simple cosmological models with shear,'' 
Mon. Not. Roy. Astron. Soc. {\bf 142} (1969) 213.  

\bibitem{koimo1}
  T.~Koivisto and D.~F.~Mota,
  ``Accelerating Cosmologies with an Anisotropic Equation of State,''
  Astrophys.\ J.\  {\bf 679} (2008) 1
  [arXiv:0707.0279 [astro-ph]].
  
  \bibitem{koimo}
  T.~Koivisto and D.~F.~Mota,
  ``Anisotropic Dark Energy: Dynamics of Background and Perturbations,''
  JCAP {\bf 0806} (2008) 018
  [arXiv:0801.3676 [astro-ph]].
  
   \bibitem{camp}
  L.~Campanelli, P.~Cea, G.~L.~Fogli and A.~Marrone,
  ``Testing the Isotropy of the Universe with Type Ia Supernovae,''
  Phys.\ Rev.\ D {\bf 83} (2011) 103503
  [arXiv:1012.5596 [astro-ph.CO]].

\bibitem{fpu}
P.~Fleury, C.~Pitrou and J.~P.~Uzan,
``Light propagation in a homogeneous and anisotropic universe,''
Phys. Rev. D \textbf{91} (2015) 043511
doi:10.1103/PhysRevD.91.043511
[arXiv:1410.8473 [gr-qc]].

\bibitem{lalu}
M.~Lachi\`eze-Rey and J.~P.~Luminet,
``Cosmic topology,''
Phys. Rept. \textbf{254} (1995), 135-214
doi:10.1016/0370-1573(94)00085-H
[arXiv:gr-qc/9605010 [gr-qc]].

\bibitem{lum}
J.~P.~Luminet,
``The Status of Cosmic Topology after Planck Data,''
Universe \textbf{2} (2016) no.1, 1
doi:10.3390/universe2010001
[arXiv:1601.03884 [astro-ph.CO]].

\bibitem{short}
G. Valent, A. Tilquin and T. Sch\"ucker, ``
The Lema{\^i}tre-Hubble diagram in axial Bianchi IX universes with comoving dust,'' Class. Quant. Grav. (2023) in press.

 \bibitem{Fubini} 
  Guido Fubini, ``Sugli spazii che ammettono un gruppo continuo de movimenti,'' Annali di Mat., ser. 3, {\bf 8} (1903) 54.

 \bibitem{berger}
 M. Berger,  ``Les vari\'et\'es riemanniennes homog\`enes normales simplement connexes \`a courbure strictement positive,'' Ann. Scuola Norm. Sup. Pisa {\bf 15} (1961)179.
 
  \bibitem{ter}
 P.~A.~Terzis,
``Faithful representations of Lie algebras and Homogeneous Spaces,''
[arXiv:1304.7894 [math.RT]].
 
\bibitem{steph}
H.~Stephani, D.~Kramer, M.~A.~H.~MacCallum, C.~Hoenselaers and E.~Herlt,
``Exact Solutions of Einstein's Field Equations,'' Cambridge University Press (2003), 
doi:10.1017/CBO9780511535185; equation (8.54), page 106.

\bibitem{ks}
R. Kantowski and R. K. Sachs,
``Some Spatially Homogeneous Anisotropic Relativistic Cosmological Models,''
 J. Math. Phys. {\bf 7} (1966) 443.
 
 \bibitem{farn}
D. L. Farnsworth, ``Some New General Relativistic Dust Metrics Possessing Isometries,''  J. Math. Phys. {\bf 8} (1967) 2315.

  \bibitem{KingEllis}
 A. R. King and G. F. R. Ellis, ``Tilted Homogeneous Cosmological Models'',
 Comm. Math. Phys. {\bf 31} (1973) 209. 
 
 \bibitem{sheik}
C.~Krishnan, R.~Mondol and M.~M.~Sheikh-Jabbari,
``Dipole Cosmology: The Copernican Paradigm Beyond FLRW,''
[arXiv:2209.14918 [astro-ph.CO]].

\bibitem{gal}
G.~Valent,
``Bianchi type II,III and V diagonal Einstein metrics re-visited,''
Gen. Rel. Grav. \textbf{41} (2009), 2433-2459
doi:10.1007/s10714-009-0774-1
[arXiv:1002.1454 [math-ph]].

 \bibitem{ak}
O.~Akarsu, E.~Di Valentino, S.~Kumar, M.~Ozyigit and S.~Sharma,
``Testing spatial curvature and anisotropic expansion on top of the $\Lambda$CDM model,''
[arXiv:2112.07807 [astro-ph.CO]].

 \bibitem{wald}
R. M. Wald, `` Asymptotic behavior of homogeneous cosmological models in the presence of a positive cosmological constant,'' Phys. Rev. D  \textbf{28} (1983) 2118.

 \bibitem{misn}
 C. W. Misner, ``Mixmaster universe,'', Phys. Rev. Lett. {\bf 22} (1969) 1071.
 
  \bibitem{bkl1}
   V. A. Belinskii, I. M. Khalatnikov, and E. M. Lifshitz,  ``Oscillatory approach to a singular point in the relativistic cosmology,''  Zh. Eksp. Teor. Fiz. {\bf 60} (1971) 1969.
  
   \bibitem{bkl2}
V.~A.~Belinskii, E.~M.~Lifshitz and I.~M.~Khalatnikov,
``On a general cosmological solution of the Einstein equations with a time singularity,''
Zh. Eksp. Teor. Fiz. \textbf{62} (1972).

\bibitem{dlh1}
P.~P.~Dechant, A.~N.~Lasenby and M.~P.~Hobson,
``An Anisotropic, non-singular early universe model leading to a realistic cosmology,''
Phys. Rev. D \textbf{79} (2009), 043524
doi:10.1103/PhysRevD.79.043524
[arXiv:0809.4335 [gr-qc]].

\bibitem{dlh2}
P.~P.~Dechant, A.~N.~Lasenby and M.~P.~Hobson,
``Cracking the Taub-NUT,''
Class. Quant. Grav. \textbf{27} (2010), 185010
doi:10.1088/0264-9381/27/18/185010
[arXiv:1007.1662 [gr-qc]].

\bibitem{gpk}
L.~Giani, O.~F.~Piattella and A.~Y.~Kamenshchik,
``Bianchi IX gravitational collapse of matter inhomogeneities,''
JCAP \textbf{03} (2022) no.03, 028
doi:10.1088/1475-7516/2022/03/028
[arXiv:2112.01869 [gr-qc]].

\bibitem{uke}
J.~P.~Uzan, U.~Kirchner and G.~F.~R.~Ellis,
``WMAP data and the curvature of space,''
Mon. Not. Roy. Astron. Soc. \textbf{344} (2003), L65
doi:10.1046/j.1365-8711.2003.07043.x
[arXiv:astro-ph/0302597 [astro-ph]].

\bibitem{div}
E.~Di Valentino, A.~Melchiorri and J.~Silk,
``Planck evidence for a closed Universe and a possible crisis for cosmology,''
Nature Astron. \textbf{4} (2019) no.2, 196-203
doi:10.1038/s41550-019-0906-9
[arXiv:1911.02087 [astro-ph.CO]].
   
\bibitem{lhtw}
L.~Liu, L.~J.~Hu, L.~Tang and Y.~Wu,
``Constraining the spatial curvature of the local Universe with deep learning,''
[arXiv:2309.11334 [astro-ph.CO]].
   
\bibitem{Ki}
D. H. King, Phys. Rev. D {\bf44} (1991) 2356.\\
Note two misprints, 
the fourth of equations (B8) should read: $y_4=r\cos\psi\sin\beta $ and (B14) should be: $dl^2\equiv{\textstyle\frac{1}{4}}S^2[(dx^1)^2+(dx^2)^2+2\cos x^1\,dx^2dx^3+(dx^3)^2]$.

\bibitem{thesis}
P.~Fleury,
``Light propagation in inhomogeneous and anisotropic cosmologies,''
[arXiv:1511.03702 [gr-qc]].

\bibitem{sign}
Note a missing $\pm$ in the corresponding equation (90) of our paper \cite{stv}. The sign in equation (95) is correct. 

 
 \bibitem{exact4}
 R.~Coquereaux and A.~Grossmann,
``Analytic Discussion of Spatially Closed Friedmann Universes With Cosmological Constant and Radiation Pressure,''
Annals Phys. \textbf{143} (1982), 296
[erratum: Annals Phys. \textbf{170} (1986) 490]
doi:10.1016/0003-4916(82)90030-6.

\bibitem{exact5}
R.~Coquereaux and A.~Grossmann,
``Large scale geometry and evolution of a universe with radiation pressure and cosmological constant,''
[arXiv:astro-ph/0101369 [astro-ph]].

\bibitem{exact6}
R.~Coquereaux,
``The history of the universe is an elliptic curve,''
Class. Quant. Grav. \textbf{32} (2015) no.11, 115013
[erratum: Class. Quant. Grav. \textbf{33} (2016) no.15, 159601]
doi:10.1088/0264-9381/32/11/115013
[arXiv:1411.2192 [gr-qc]].

\bibitem{edw} 
 D. Edwards, 
``Exact expressions for the properties of the zero-pressure Friedmann models,'' Mon.  R. Astron. Soc.  \textbf{159} (1972) 51-66.

\bibitem{bf} 
P. F. Byrd and M. D. Friedman, ``Handbook of Elliptic Integrals for Engineers and Scientists,'' 2nd edition,  Springer-Verlag (1971), page 135.

\bibitem{ww}
 E. T. Whittaker and G. N. Watson, ``A Course of Modern Analysis,\\ 4th edition, Cambridge University Press (1986), page 464.

\bibitem{jla}  
M.~Betoule \textit{et al.} [SDSS],
``Improved cosmological constraints from a joint analysis of the SDSS-II and SNLS supernova samples,''
Astron. Astrophys. \textbf{568} (2014) A22
doi:10.1051/0004-6361/201423413
[arXiv:1401.4064 [astro-ph.CO]].

\bibitem{jwst}
J.~Lu, L.~Wang, X.~Chen, D.~Rubin, S.~Perlmutter, D.~Baade, J.~Mould, J.~Vinko, E.~Reg\H{o}s and A.~M.~Koekemoer,
``Constraints on Cosmological Parameters with a Sample of Type Ia Supernovae from JWST,''
Astrophys. J. \textbf{941} (2022) no.1, 71
doi:10.3847/1538-4357/ac9f49
[arXiv:2210.00746 [astro-ph.CO]].
\bibitem{csst}
S.~Y.~Li, Y.~L.~Li, T.~Zhang, J.~Vinko, E.~Regos, X.~Wang, G.~Xi and H.~Zhan,
``Forecast of cosmological constraints with type Ia supernovae from the Chinese Space Station Telescope,''
Sci. China Phys. Mech. Astron. \textbf{66} (2023) no.2, 229511
doi:10.1007/s11433-022-2018-0
[arXiv:2210.05450 [astro-ph.CO]].

\bibitem{tl}
  O.  Titov and S. Lambert,
  ``On the VLBI measurement of the Solar System acceleration'', Astron.\ Astrophys.\  {\bf 559} (2013) A95
  [arXiv:1310.2723 [astro-ph.IM]].

\bibitem{cea22}
P.~K.~Aluri, P.~Cea, P.~Chingangbam, M.~C.~Chu, R.~G.~Clowes, D.~Hutsem\'ekers, J.~P.~Kochappan, A.~M.~Lopez, L.~Liu and N.~C.~M.~Martens, \textit{et al.}
``Is the observable Universe consistent with the cosmological principle?,''
Class. Quant. Grav. \textbf{40} (2023) no.9, 094001
doi:10.1088/1361-6382/acbefc
[arXiv:2207.05765 [astro-ph.CO]].

\bibitem{hall}
G.~S.~Hall,
``On the theory of Killing orbits in space-time,''
Class. Quant. Grav. \textbf{20} (2003), 4067
doi:10.1088/0264-9381/20/18/313
[arXiv:gr-qc/0310074 [gr-qc]];\\
 Theorem 8.

\end{thebibliography}
\end{document}